\documentclass[a4paper,11pt]{article}
\usepackage{jcappub}

\usepackage{ulem}
\usepackage{booktabs}
\usepackage[english]{babel}
\usepackage{amsmath,amssymb,amsbsy,amstext, amsthm, simplewick}
\usepackage{hyperref}
\usepackage{graphicx}
\usepackage{amsfonts}
\usepackage{amssymb}
  \usepackage{framed}
\usepackage{enumitem}
\usepackage{soul}
\usepackage{orcidlink}
\usepackage{upgreek}
 \usepackage{exscale,relsize}
 \usepackage[makeroom]{cancel}
\usepackage{soul}
\usepackage{bbold}
\usepackage{lipsum}
\usepackage{mdframed}
\usepackage{mathtools}
\usepackage[dvipsnames]{xcolor}
\usepackage{tikz}
\usepackage{tikz-cd}
\usepackage[export]{adjustbox}
\usepackage{dsfont}
\usepackage[mathscr]{eucal}
\usepackage{cleveref}
\usepackage{comment}
\usepackage{wrapfig}
\usepackage{booktabs,multirow}

 \newcommand{\circled}[2][]{%
  \tikz[baseline=(char.base)]{%
    \node[shape = circle, draw, inner sep = 1pt]
    (char) {\phantom{\ifblank{#1}{#2}{#1}}};%
    \node at (char.center) {\makebox[0pt][c]{#2}};}}
\robustify{\circled}

\newcounter{qnumber}

\newcounter{qnumber2}
\definecolor{c2}{rgb}{0.45, 0.876, 0.2}

\usepackage{colortbl}
\definecolor{lightgreen}{cmyk}{0.2, 0, 0.2, 0.2}
\definecolor{lightgray}{cmyk}{0.1,0.2,0,0.1}
\definecolor{lightgray2}{cmyk}{0.1,0.1,0,0.1}

\makeatletter
\newlength{\apb@width}
\newcommand{\autoparbox}[2][c]{\settowidth{\apb@width}{#2}\parbox[#1]{\apb@width}{#2}}

\newcommand{\f}{\varphi}
\newcommand{\Cen}[2]{%
  \ifmeasuring@
    #2%
  \else
    \makebox[\ifcase\expandafter #1\maxcolumn@widths\fi]{$\displaystyle#2$}%
  \fi
}
\makeatother

\newcommand{\beq}{\begin{equation}\begin{aligned}}
\newcommand{\eeq}{\end{aligned}\end{equation}}

\RequirePackage{amsmath}
\RequirePackage{amssymb}
\RequirePackage{epsfig}
\RequirePackage{graphicx}
\RequirePackage[numbers,sort&compress]{natbib}
\RequirePackage[colorlinks=true
  ,urlcolor=blue
  ,anchorcolor=blue
  ,citecolor=blue
  ,filecolor=blue
  ,linkcolor=blue
  ,menucolor=blue
  ,pagecolor=blue
  ,linktocpage=true
  ,pdfproducer=medialab
  ,pdfa=true
]{hyperref}

\def\beq{\begin{equation}}
\def\eeq{\end{equation}}

\def\Beq{\begin{equation}\begin{aligned}}
\def\Eeq{\end{aligned}\end{equation}}

\def\bea{\begin{eqnarray}}
\def\eea{\end{eqnarray}}

\def\beq{\begin{equation}}
\def\eeq{\end{equation}}
\def\bea{\begin{eqnarray}}
\def\eea{\end{eqnarray}}

\newcommand\e{\varepsilon}

\DeclareRobustCommand{\SkipTocEntry}[4]{}
\DeclareSymbolFont{extraup}{U}{zavm}{m}{n}
\DeclareMathSymbol{\varheart}{\mathalpha}{extraup}{86}
\DeclareMathSymbol{\vardiamond}{\mathalpha}{extraup}{87}

\definecolor{darkgreen}{rgb}{0,0.5,0}

\crefname{equation}{Eq.}{Eqs.}
\Crefname{equation}{Equation}{Equations}

\title{\huge Gravitational Waves from Long Strings and Loops}

\author[a]{
Christos Litos\ \orcidlink{0009-0009-8823-6264},}
\author[a, b]{Sarunas Verner\ \orcidlink{0000-0003-4870-0826},}
\author[a]{Wei Xue\ \orcidlink{XXXX-XXXX-XXXX-XXXX},} 
\author[a, c, d]{Fengwei Yang\ \orcidlink{0000-0001-9873-6259}}

\affiliation[a]{Institute for Fundamental Theory, Physics Department, University of Florida,\\ Gainesville, FL 32611, USA}

\affiliation[b] {Kavli Institute for Cosmological Physics, 
University of Chicago, 5640 South Ellis Ave., Chicago, IL 60637, USA}

\affiliation[c]{Department of Physics and Astronomy,
University of Notre Dame, South Bend, IN 46556, USA}

\affiliation[d]{Johannes Gutenberg-Universit\"{a}t Mainz, 55099 Mainz, Germany}

\emailAdd{c.litos@ufl.edu}
\emailAdd{verner@uchicago.edu}
\emailAdd{weixue@ufl.edu}
\emailAdd{fyang@uni-mainz.de}

\abstract{We compute the gravitational wave spectrum produced by a global cosmic string network in the scaling regime. The spectrum naturally divides at the string correlation length into infrared and ultraviolet parts. In the infrared, we derive the spectrum analytically from the unequal-time correlator of the string stress-energy tensor within the unconnected segment model, which we generalize for the first time to string loops. The loop contribution can be comparable to that of long strings, depending on the loop evolution parameters. In the ultraviolet, where structure below the correlation length dominates the source, we develop a data-driven method that connects this correlator formalism to the instantaneous radiation power spectrum measured in existing lattice simulations. The predicted spectrum rises with frequency as $\Omega_{\rm GW}\propto k$ and $k^{3}$ in the infrared, and flattens into a plateau with a mild logarithmic tilt in the ultraviolet. The turnover frequency between the two is set by the mass of the Goldstone boson, the axion. Confronting the predicted spectrum with current data and projected sensitivities, we map out current constraints and future probes in the plane of the symmetry-breaking scale~$f_a$ and the axion mass~$m_a$.
}

\begin{document}
\maketitle
\flushbottom
\newpage
\section{Introduction}
\label{sec:introduction}

The direct detection of gravitational waves (GWs) by the LIGO-Virgo collaboration in 2015~\cite{LIGOScientific:2016aoc} marked the dawn of GW astronomy, opening an unprecedented window into the dynamics of compact objects and, potentially, the physics of the early universe. While initial detections focused on transient events such as binary black hole and neutron star mergers~\cite{LIGOScientific:2017vwq}, attention has increasingly turned to the stochastic gravitational wave background (SGWB), a superposition of unresolved GW signals permeating the cosmos. In 2023, pulsar timing array (PTA) collaborations including NANOGrav~\cite{NANOGrav:2023gor,NANOGrav:2023hvm}, EPTA~\cite{EPTA:2023fyk}, PPTA~\cite{Reardon:2023gzh}, and CPTA~\cite{Xu:2023wog} reported evidence for a stochastic process in the nanohertz frequency band exhibiting the Hellings-Downs angular correlations expected of an SGWB. While the signal may originate from a population of supermassive black hole binaries~\cite{NANOGrav:2023hfp}, cosmic strings remain a viable and intriguing alternative~\cite{NANOGrav:2023hvm, Auclair:2019wcv, Ellis:2023tsl}.

Cosmic strings are one-dimensional topological defects that form during symmetry-breaking phase transitions in the early universe~\cite{Kibble:1976sj, Vilenkin:2000jqa, Hindmarsh:1994re}. When a group $G$ is spontaneously broken to a subgroup $H$, the topology of the vacuum manifold $\mathcal{M} = G/H$ determines which defects can form: a nontrivial first homotopy group, $\pi_1(\mathcal{M}) \neq 1$, signals the existence of stable string solutions~\cite{Kibble:1976sj}. The prototypical example is a complex scalar field with a Mexican hat potential that undergoes spontaneous symmetry breaking via a second-order phase transition. After the phase transition, the phase of the scalar field varies randomly between uncorrelated regions, and strings inevitably form, with a characteristic separation set by the correlation length~\cite{Kibble:1980mv,Zurek:1985qw}. Such networks are predicted in a wide range of beyond Standard Model scenarios, including grand unified theories~\cite{Jeannerot:2003qv}, superstring theory~\cite{Copeland:2003bj}, and axion physics~\cite{Kim:1986ax,Marsh:2015xka}.

The dynamics of cosmic string networks are rich and complex, involving the formation and propagation of small-scale structure such as kinks and cusps, the self-intersection of long strings to form loops, and the gradual transfer of energy from the infinite-string network to subhorizon loops and radiation~\cite{Vilenkin:2000jqa,Copeland:2009ga}. Full numerical simulations of string networks are computationally demanding, with only a limited number of $e$-folds of evolution achievable~\cite{Blanco-Pillado:2013qja, Hindmarsh:2017qff}. However, semi-analytic approaches based on the velocity-dependent one-scale (VOS) model~\cite{Martins:1996jp, Martins:2000cs} have proven remarkably successful at capturing the macroscopic network evolution. The VOS model tracks the root-mean-square (RMS) velocity $v$ and the correlation length $\ell_c = \xi \eta$ (where $\xi$ is a dimensionless parameter and $\eta$ is conformal time) through a set of coupled ordinary differential equations. A key prediction is the existence of a \textit{scaling regime}: an attractor solution in which $\xi$ and $v$ approach constant values, and the network maintains a fixed number of strings per Hubble volume throughout cosmic history~\cite{Martins:2000cs,Sousa:2013aaa}.

If the broken symmetry is gauged (local), one obtains Nielsen-Olesen strings~\cite{Nielsen:1973cs}, whose energy density is confined to a finite-width core set by the inverse masses of the scalar and gauge fields,
\begin{equation}
\delta \; = \; \max\{m_{\phi}^{-1},\,m_A^{-1}\}, \qquad
m_{\phi} = \sqrt{\lambda/2}\,f_a,\quad m_A =e\,f_a,
\label{eq:corewidth}
\end{equation}
so parametrically $\delta\sim f_a^{-1}$ up to couplings~\cite{Nielsen:1973cs,Vilenkin:2000jqa}. If the symmetry is global, the string is instead accompanied by a long-range massless Goldstone field, and the energy per unit length acquires a logarithmic infrared enhancement,
\begin{equation}
\mu_{\rm global} \; \simeq \; \pi f_a^2 \ln\!\left(\frac{L}{\delta}\right),
\label{eq:muglobal}
\end{equation}
where $L$ is an infrared cutoff (e.g. the correlation length or the horizon scale), and the associated stress extends to large distances \cite{Hindmarsh:1994re, Vilenkin:2000jqa}.

The nature of the underlying symmetry, global or gauged, has profound implications for the string network evolution and decay. \textit{Global} (or axion) strings arise from the breaking of a global $\mathrm{U}(1)$ symmetry, such as the Peccei-Quinn (PQ) symmetry invoked to solve the strong CP problem~\cite{Peccei:1977hh,Weinberg:1977ma, Wilczek:1977pj}. These strings store energy not only in their cores but also in long-range Goldstone-boson gradients, leading to a logarithmically enhanced tension~\cite{Vilenkin:1984ib, Davis:1986xc}. When the associated pseudo-Nambu-Goldstone boson (the axion) acquires a mass (typically when $H \sim m_a$) domain walls form at string endpoints, causing the network to collapse and decay into axions, radial modes, and gravitational waves~\cite{Vilenkin:1982ks,Chang:1998tb,Hiramatsu:2012gg,Gorghetto:2018myk}. \textit{Gauge} (or local) strings, by contrast, have their energy confined to a microscopic core and can persist indefinitely. Their primary decay channel is through the formation of loops, which oscillate and radiate energy predominantly into gravitational waves over many oscillation periods~\cite{Vachaspati:1984gt,Burden:1985md,Blanco-Pillado:2017oxo}.

The GW signal from cosmic strings has been studied extensively, both analytically and numerically~\cite{Vilenkin:1981bx,Hogan:1984is,Vachaspati:1984gt,Caldwell:1991jj,Damour:2001bk,Damour:2004kw,Siemens:2006yp,Olmez:2010bi, Sanidas:2012ee, Blanco-Pillado:2017rnf, Auclair:2019wcv, Gouttenoire:2019kij}. For Nambu-Goto strings (the infinitely thin approximation valid for gauge strings), the dominant contribution at high frequencies is often attributed to loops, while long strings contribute primarily at lower frequencies~\cite{Caldwell:1991jj,Figueroa:2012kw,Sousa:2020sxs}. Global strings present additional complications: the scale-dependent effective tension introduces logarithmic corrections to the spectrum, and the eventual network collapse sets a characteristic turnover frequency~\cite{Chang:1998tb,Gorghetto:2021fsn,Buschmann:2021sdq,Chang:2021afa}. Lattice simulations~\cite{Hindmarsh:2017qff,Hindmarsh:2019csc,Gorghetto:2020qws,Buschmann:2021sdq} and the unconnected segment model (USM)~\cite{Pen:1997ae,Pogosian:1999np,Avgoustidis:2012gb} have been used to compute unequal-time correlators (UETCs) of the string stress-energy tensor, which serve as the source terms for cosmological perturbations. The USM, in particular, has been applied extensively to CMB anisotropies but not to the stochastic GW background. Moreover, no existing calculation provides the complete GW spectrum, i.e., from the infrared regime set by the correlation length to the ultraviolet
regime governed by sub-correlation-length structure, within a single unified framework.

In this work, we address these gaps by providing a semi-analytical calculation of the GW spectrum from a scaling global string network, including long strings and loops. We construct the full UETC for long strings within the USM framework, and extend it for the first time to incorporate the loop contribution.
The scaling evolutions of long strings and loops are determined by the VOS model. 
We then evolve the source governed by the UETCs using retarded Green's functions for tensor perturbations in both radiation- and matter-dominated backgrounds, obtaining 
the GW spectral density. The calculation naturally separates into two regimes: for $k\xi\eta\lesssim 1$, the spectrum is fully captured by the USM, while for $k\xi\eta\gtrsim 1$, the straight-segment approximation breaks down and we develop a data-driven approach that uses the instantaneous radiation power spectrum from lattice simulations as input. This formalism can also be applied to scaling gauge string networks, but possible modifications are needed to properly include the IR contribution from the long-lived gauge string loops. 

Our analysis yields several results. For gauge strings, the IR spectrum of long strings reproduces Nambu-Goto simulations~\cite{CamargoNevesdaCunha:2022mvg}, validating the USM in this regime. For global strings, the logarithmic running of the effective tension enhances the long-string IR spectrum by two powers of the logarithm and the data-driven UV spectrum by three, and the loop contribution is comparable to that of long strings in the IR. Confronting the predicted spectra with current and projected observational sensitivities across more than 20 decades in frequency, we derive constraints on the symmetry-breaking scale~$f_a$ and the axion mass~$m_a$.

The remainder of this paper is organized as follows. Section~\ref{sec:cs} presents the theoretical framework for describing perturbations sourced by cosmic string networks, including four key ingredients: the energy-momentum tensor of cosmic string networks, the VOS model, the USM, and the UETCs. Section~\ref{sec:gwstrings} develops the formalism for computing the GW spectrum from the UETCs with the Green's function approach, where the analytical IR GW spectra from long-string and loop contributions are obtained, and the data-driven method for computing the UV GW spectra is introduced. Section~\ref{sec:GWSpectra} validates the USM against Nambu-Goto simulations for gauge strings, presents the GW spectra from a scaling global string network across IR ($k\lesssim\ell_c^{-1}$) and UV ($k\gtrsim\ell_c^{-1}$) scales, compares them with the sensitivity of current and future observatories, and derives constraints on the axion parameter space. We conclude in Section~\ref{sec:Conclusions}. Appendix~\ref{app:UETC} contains the full analytical expressions for the UETCs. Appendix~\ref{app:greenfunctions} includes the derivation of the Green's functions. Appendix~\ref{app:ModeMatching} discusses the general expressions used to compute the GW intensity.

\section{Cosmic String Network}
\label{sec:cs}

In this section, we present the key theoretical ingredients required 
to describe perturbations sourced by cosmic string networks, focusing on global strings. Our framework consists of four main components:
\begin{itemize}
\item the cosmic string energy-momentum tensor;

\item the velocity-dependent one-scale (VOS) model, which describes
  the macroscopic evolution of the long-string correlation length and
  root-mean-square velocity, together with the production rate of
  string loops;

\item the unconnected segment model (USM), extended to include loop contributions in the infrared;

\item the unequal-time correlation functions (UETCs).
\end{itemize}

Several features distinguish our study from previous works. First, we
generalize the gauge-string USM framework to global strings by
computing their unequal-time correlation functions and incorporating
shrinking segments for loop dynamics. Second, compared with existing
semi-analytical and numerical studies~\cite{CamargoNevesdaCunha:2022mvg,Gorghetto:2021fsn}, our approach yields a
model-independent infrared (IR) spectral shape while incorporating a
data-driven ultraviolet (UV) spectrum. This unified treatment
consistently captures both the large-scale and small-scale features of
the string network.

\subsection{Global String Energy-Momentum Tensor}

Here we present the energy-momentum tensor for straight global string segments and their Fourier transform. 

We consider the case in which global strings arise from the spontaneous breaking of a global $\rm U(1)$ symmetry by a complex scalar field $\varphi(x)$, which acquires a vacuum expectation value (vev) of $\langle \varphi \rangle = f_a / \sqrt{2}$.
The action for the field $\varphi$ is given by  
\begin{equation}
   S  = \int d^4x \sqrt{-g} \left(    \partial_\mu  \varphi \partial^\mu \varphi^\dagger  - V ( \varphi) \right) \, , 
\end{equation}
where the potential $V(\varphi) = \frac{\lambda}{4}  \left( |\varphi|^2 - f_a^2/2 \right)^2$. 
The background spacetime is taken to be the Friedmann-Lema\^itre-Robertson-Walker (FLRW) metric for a flat universe,
\begin{equation}
    ds^2= a^2(\eta) \left( d\eta^2 - \delta_{ij} dx^idx^j \right) \, .
\end{equation}
The conformal time interval $d\eta$ can be converted to the physical time interval via $dt = a(\eta ) d\eta$. Throughout this analysis, we use both conformal and physical time descriptions as convenient.

We analyze a straight string segment of length $\ell_c$, representing the macroscopic correlation length of the string network. 
Within this correlation length, the string profile can be approximated as straight, whereas at larger scales, 
the string trajectories follow random walks.
There are two primary reasons a straight-string approximation is well-suited for global strings. 
First, sub-horizon small-scale structure (such as wiggliness) is efficiently smoothed out by the rapid radiation of axions, 
the Goldstone bosons associated with the global symmetry breaking. For gauge strings, which lack this efficient radiative smoothing channel, 
wiggliness is instead typically parameterized by introducing a modifying factor $\alpha_{\rm w}$ to the energy-momentum tensor 
\cite{Carter:1990nb,Vilenkin:1990mz}. Second, this assumption is valid in the infrared (IR) regime, where the characteristic wavelength of 
interest is much larger than the string correlation length, allowing us to safely neglect small-scale string dynamics.

For a static string oriented strictly along the $z$-axis, the scalar field profile evaluated in the transverse plane $(r, \phi)$ takes the form
\begin{equation}
   \varphi (r, \phi)  = \frac{1}{\sqrt{2} } f_a e^{i \phi } \, , 
\end{equation}
valid at distances much larger than the string core size ($r \gg m_\varphi^{-1}$). 
The core size is estimated by the inverse scalar mass $\delta \sim 1/m_\varphi$, where $m_\varphi = \sqrt{\lambda/2}\, f_a$.
Accordingly, the non-zero energy momentum tensor for the straight string is given as 
\begin{eqnarray}
   T_{00} ({\bf x}, \eta ) &=&  - T_{33} ( {\bf x}, \eta )  = \frac{f_a^2} {2} \frac{ 1 } {r^2}    \, , 
   \\
   T_{\hat{i} \hat{j}} ({\bf x}, \eta )  &=&    - \frac{f_a^2} {2} \frac{ 1 } {r^2}  \delta_{ \hat{i} \hat{j}}  
      + f_a^2 \epsilon_{  \hat{i} \hat{k}} \epsilon_{  \hat{j} \hat{\ell }} \frac{   x_{\hat{k} }  x_{\hat{\ell} } } { r^4} 
      \, , \quad
      {\rm where }   \,\,\, \hat{i} ,  \hat{j},  \hat{k},  \hat{ \ell } = 1 , 2 \, , 
\end{eqnarray}
and $\epsilon_{  \hat{i} \hat{j}}$ represents the two-dimensional Levi-Civita symbol. 
In this derivation, we have neglected the FLRW metric perturbations as well as  the string core contribution. 
The string core contribution can be captured by the Nambu-Goto string solutions, but due to the logarithmic enhancement in the string tension 
from the outside region, the core contribution is negligible. 

Next, we calculate the Fourier transform of $T_{\mu \nu } ({\bf x}, \eta )$ to obtain its momentum-space representation $T_{\mu \nu} ( {\bf k}, \eta)$. 
Before addressing segments with arbitrary positions, orientations, and velocities, computing the transformation of a static straight segment 
centered at the origin and aligned along the $z$-axis will help us further simplify $T_{\mu \nu } ({\bf x}, \eta )$. The 00-component of the energy-momentum tensor 
in Fourier space is found to be 
\begin{equation}
   \begin{split}
   T_{00}^{(z)}  (  {\bf k}, \eta ) &= \int d^3 {\bf x}\, e^{- i {\bf k } \cdot {\bf x } }  T_{00} ({\bf x}, \eta )  
          = \pi f_a^2 \frac{  \sin ( k_{\parallel} \ell_{c} / 2) } { k_{\parallel} /2} 
      \int \frac{{\rm d} r}{r}  J_0 ( k_{\perp} r )  
      \\
      & \simeq
       \pi f_a^2 \frac{  \sin ( k_{\parallel} \ell_{c} / 2) } { k_{\parallel} /2} 
      \times
   \begin{cases}
     \ln (  \ell_c a (\eta ) / \delta  ),  & \quad   k_{\perp}  \ell_c \ll    1 \\
     -  \ln (  k_{\perp}  \, \delta  / a (\eta ) ),   & \quad k_{\perp}  \ell_c \gg    1 
\end{cases},
   \end{split}
\label{eq:T00_z}
\end{equation}
where $k_{\parallel}$ represents the magnitude of the momentum component parallel to the string's orientation, and $k_{\perp}$ is the magnitude of the perpendicular momentum component. Here, we have approximated the length of an uncorrelated string segment using the correlation length $\ell_{c}$. In particular, when the string network reaches the scaling regime, its comoving correlation length is characterized by $\ell_{c} = \xi \eta$, where the scaling parameter $\xi \lesssim {\cal O}(1)$ is calibrated from numerical simulations.

Following a similar procedure for the transverse spatial components $T_{\hat{i} \hat{j}} ({\bf k}, \eta)$, we find that their Fourier transforms do not 
exhibit the same logarithmic enhancement as present in $T_{00} ({\bf k}, \eta) = - T_{33} ({\bf k}, \eta)$. Consequently, we may further simplify the tensor 
formulation by entirely neglecting $T_{\hat{i} \hat{j}} ({\bf k}, \eta)$, yielding $T_{\mu \nu} ({\bf k}, \eta) \simeq {\rm diag}\left( T_{00} ({\bf k}, \eta), 0 , 0, - T_{00} ({\bf k}, \eta) \right)$.

The preceding analysis was performed in the rest frame of a static string aligned along the $z$-axis. 
To construct a segment characterized by a general orientation and velocity, 
we apply a sequence of Lorentz transformations to the string while choosing the momentum vector ${\bf k}$ 
along the $z$-direction. If the string segment is centered at ${\bf x}_0$ and its orientation lies along 
$\hat{\bf x} =(\sin \theta \cos \phi, \sin \theta \sin \phi, \cos \theta)$, we assume that its velocity is transverse to the string direction.
We neglect any velocity component parallel to the string, since longitudinal motion corresponds to a reparametrization of the string worldsheet and is unphysical.
To obtain the general configuration, we apply the following 
composite Lorentz transformation, $\Lambda  = \Lambda_\phi \Lambda_\theta \Lambda_\beta $, to our static $z$-aligned string solution, followed by a translation of the segment midpoint to ${\bf x}_0$, which in Fourier space contributes the phase factor $e^{-i{\bf k}\cdot{\bf x}_0}$.
Here, $\Lambda_\beta$ boosts the string in the $(\cos \psi, \sin \psi, 0)$ direction; $\Lambda_\theta$ rotates the string's orientation within the $(x,z)$ plane by a polar angle $\theta$; and $\Lambda_\phi$ governs an azimuthal rotation in the $(x,y)$ plane by angle $\phi$. 
Consequently, the energy-momentum tensor of the general configuration is obtained as $T_{\mu \nu} \to \Lambda T \Lambda^T$.

Therefore, for a straight string segment oriented along $\hat{\bf x} = (\sin \theta \cos \phi, \sin \theta \sin \phi, \cos \theta)$ with momentum evaluated along the $z$-direction, the energy-momentum tensor in Fourier space can be explicitly written as 
\begin{equation}
   T_{00}  (  {\bf k}, \eta ) 
       \simeq
       \frac{\pi f_a^2 }{\sqrt{1 - v^2}} \frac{  \sin ( k \ell_{c} \cos \theta / 2) } { k \cos \theta  /2}\, e^{-i{\bf k}\cdot{\bf x}_0}\, e^{-ikv\eta\cos\psi\sin\theta}
      \times
   \begin{cases}
     \ln (  \ell_c a (\eta ) / \delta  ),  & \quad   k \ell_c \sin \theta   \ll    1 \\
     -  \ln (  k\sin\theta   \, \delta / a (\eta ) ),   & \quad k \ell_c \sin \theta   \gg    1 
\end{cases},
\label{eq:T00}
\end{equation}
where $k\equiv|{\bf k}|$. The spatial components satisfy the relation
\begin{equation}
\label{eq:T_ij}
    T_{ij}({\bf k},\eta) \; = \; T_{00}({\bf k},\eta)\left[v^2 \hat{v}^i \hat{v}^j-(1-v^2)\hat{x}^i \hat{x}^j\right], ~~i,j \; = \; 1,2,3,
\end{equation}
with $\hat{v}$ given in \cref{eq:B3}. In the IR regime, we observe that the global string's energy-momentum tensor 
reduces to that of a local string, 
\begin{equation}   
   T_{00}^{\rm (IR)} (  {\bf k}, \eta ) = \frac{\mu_{\rm eff} ( \eta)  }{\sqrt{1 - v^2}} \frac{  \sin ( k \ell_{c} \cos \theta / 2) } { k \cos \theta  /2}\, e^{-i{\bf k}\cdot{\bf x}_0}\, e^{-ikv\eta\cos\psi\sin\theta}
\end{equation}   
provided we identify its effective tension as 
\begin{equation}
\label{eq:mu_eff}
\mu_{\rm eff} ( \eta) =   \pi f_a^2  \ln (  \ell_c a (\eta ) / \delta  ) \, .
\end{equation}
Thus, the energy-momentum tensor for a global string mirrors that of a gauge string, but exceeds it by a $k$-independent logarithmic enhancement encoded in the effective tension. In the short-wavelength limit, by contrast, the IR cutoff of the logarithm is set by the wavelength of the mode rather than by the correlation length.

\subsection{Velocity-Dependent One-Scale Model}

The macroscopic evolution of a cosmic string network is governed by the velocity-dependent one-scale (VOS) model~\cite{Kibble:1984hp,Austin:1993rg,Martins:1996jp, Martins:2000cs}.
This model tracks two averaged macroscopic quantities: the correlation length, denoted by $L_c$ for the physical length and $\ell_c = L_c/a$ for the comoving length, and 
the root-mean-square (RMS) velocity $v$, defined as $v \equiv \sqrt{\langle|d\mathbf{x}/d\eta|^2\rangle}$. 
As the string network evolves, it asymptotically approaches a scaling solution 
in which the correlation length scales proportionally to the Hubble radius, such that $\ell_c = \xi \eta$ where $\xi$ approaches a constant value. 
The VOS model accurately reproduces the large-scale properties of scaling gauge string networks found in numerical simulations~\cite{Bennett:1989yp,Allen:1990tv}. 
For global strings, however, the scaling behavior is modified. Because the effective tension of a global string grows logarithmically with time, numerical simulations~\cite{Gorghetto:2021fsn} indicate a logarithmic violation of exact scaling: the number of long strings per Hubble volume grows as $N_{\rm str}\simeq c_1\ln(\ell_c a/\delta)+c_0$, so that $\xi$ decreases slowly over time rather than remaining strictly constant. In this work, we neglect this slow evolution and adopt constant scaling parameters. 

The evolution of the long-string energy density can be deduced from the VOS equation for the string correlation length. By taking $\rho_{\rm str} = \mu / L_c^2$, the energy density evolution is given by
\begin{equation}
\label{eq:rho_str_evo}
    \frac{d\rho_{\rm str}}{dt} + 2H\rho_{\rm str}(1+v^2) = - \tilde{c} v \frac{\mu L_c}{L_c^4},
\end{equation}
where $\tilde{c}$ is the loop-chopping efficiency parameter. We neglect the additional energy-loss term from Nambu-Goldstone boson radiation, since its corrections to the correlation length and RMS velocity are suppressed by the large logarithm $\ln(\ell_c a/\delta)$. See Ref.~\cite{Martins:2018dqg} for the full VOS equations of global strings.

Combined with the companion equation governing the evolution of $v$~\cite{Martins:1996jp,Martins:2000cs}, the VOS system admits analytical scaling solutions in pure radiation-dominated (RD) and matter-dominated (MD) backgrounds.
The asymptotic scaling values of $\xi$ and $v$ are calibrated to Nambu-Goto simulations~\cite{Martins:2000cs} for gauge strings, and to field-theory simulations~\cite{Martins:2018dqg} 
for global strings. These parameters are collected in Table~\ref{tab:VOS_scaling}. 

\begin{table}[ht]
  \centering
  \renewcommand{\arraystretch}{1.4}
  \begin{tabular}{l@{\hskip 1.2em}c@{\hskip 1.8em}cc@{\hskip 1.8em}cc}
    \toprule
    \multirow{2}{*}{\textbf{String type}}
    & \multirow{2}{*}{$\boldsymbol{\tilde{c}}$}
    & \multicolumn{2}{c@{\hskip 1.8em}}{$\boldsymbol{\xi}$}
    & \multicolumn{2}{c}{$\boldsymbol{v}$} \\
    \cmidrule(lr){3-4}\cmidrule(lr){5-6}
    & & RD & MD & RD & MD \\
    \midrule
    Gauge (Nambu-Goto) & $0.23$ & $0.13$ & $0.20$ & $0.65$ & $0.58$ \\[3pt]
    Global              & $0.66$ & $0.24$ & $0.17$ & $0.51$ & $0.37$ \\
    \bottomrule
  \end{tabular}
    \caption{VOS model loop-chopping efficiency~$\tilde{c}$ and
    asymptotic scaling solutions for gauge (Nambu-Goto)~\cite{Martins:2000cs} and global
    string networks~\cite{Martins:2018dqg} in the radiation-dominated (RD) and
    matter-dominated (MD) eras, calibrated to
    simulations. Note that the scaling solutions of global string networks in MD are derived from the VOS model using the same VOS parameters as in RD, consistent with the results in \cite{Chang:2021afa}.
}
  \label{tab:VOS_scaling}
\end{table}

String loops are produced when long strings intercommute and self-intersect. The number density of string loops, $n_{\rm loop}$, satisfies the continuity equation
\begin{equation}
\label{eq:loop_form_rate}
     \frac{d n_{\rm loop}}{dt} + 3H n_{\rm loop} = 
     \frac{g \tilde{c}v} {L_c^3 L_{\rm loop}}, 
\end{equation}
where $L_{\rm loop}=a\,\ell_{\rm loop}$ is the physical loop length at formation, and $g\simeq\sqrt{1-v^2}$ is a correction factor accounting for the bulk velocity of loops at formation: the kinetic energy of a loop redshifts away, so only a fraction $g$ of the energy transferred from long strings ends up as loop rest energy. Note that \cref{eq:loop_form_rate} does not include the decay of string loops through Nambu-Goldstone boson or gravitational wave radiation, which is modeled separately as the shrinkage of loops in \cref{eq:ell_loop}.
To estimate the loop contribution to the gravitational wave signal, we assume that loops formed at conformal time $\eta$ are characterized by a typical comoving length $\ell_{\rm loop} = \alpha \eta$. In our analysis, we adopt $\alpha=0.1$. 

\subsection{Unconnected Segment Model and String Loops}
\label{sec:USLM}

\begin{figure*}[t!]
        \centering
         \includegraphics[scale = 0.5]{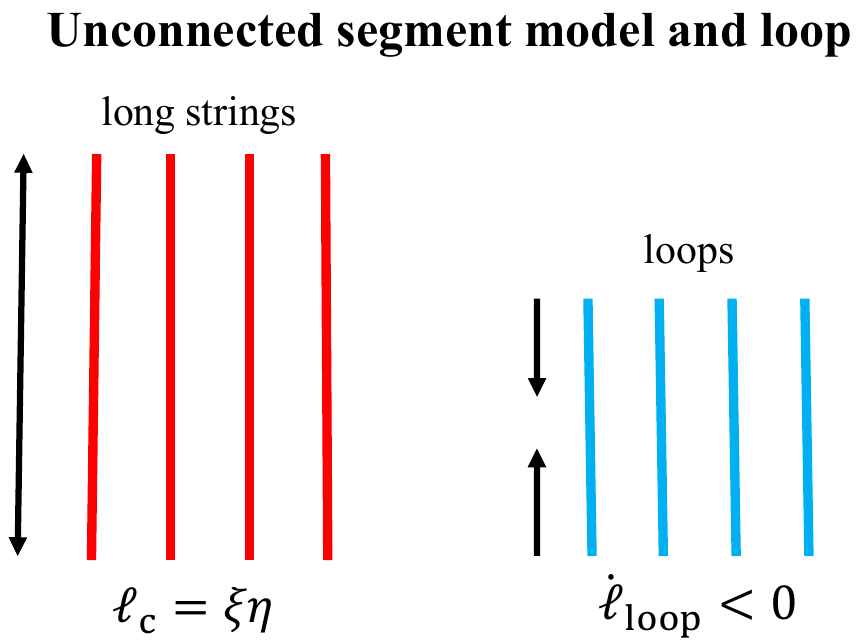}
    \caption{A schematic plot of the unconnected string and loop segments in the USM.} 
    \label{fig:uslm}
\end{figure*}

In this subsection, we present the unconnected segment model (USM)~\cite{Albrecht:1997mz,Pogosian:1999np} for global long strings, 
and extend this analysis to incorporate string loops. 
The long-string dynamics are described by the USM, while the loop sector is modeled via a distribution of shrinking string loops, as illustrated in Fig.~\ref{fig:uslm}.
Both components contribute to the IR spectrum of the stochastic gravitational wave background. As we show in Sec.~\ref{sec:gwstrings}, the two contributions turn out to be comparable. 

The USM is an analytical framework for computing two-point correlation functions in a scaling string network. 
It has been extensively used to derive large-scale scalar, vector, and tensor perturbation spectra in the CMB sourced by Nambu-Goto strings~\cite{Albrecht:1997mz,Pogosian:1999np,Avgoustidis:2012gb,Lazanu:2014xxa, Charnock:2016nzm,Rybak:2021scp,Silva:2023diq,Raidal:2026cpb,Caloni:2026dyu}, and more recently, the radiation power spectrum from global strings~\cite{Correia:2025nns}. 

In the USM, the long-string network is modeled as an ensemble of uncorrelated, randomly oriented straight segments, 
each of comoving length $\ell=\xi\eta$, moving at a velocity $v$ transverse to its orientation. 
In the long-wavelength regime, $k \lesssim 1/\ell$, the small-scale wiggliness of the strings is irrelevant, and each segment is effectively featureless. Therefore, the USM accurately captures the IR two-point statistics of the long-string network.
In this model, the comoving number density of string segments, $\tilde{n}_{\rm str} = 1/(\xi\eta)^3$, decreases with time, while the length of each segment increases with time. 

A natural question is how much string loops contribute to the gravitational perturbations. To address this, we extend the USM framework to include string loops. Similar to the long-string treatment, we focus on the long-wavelength regime, $k \lesssim 1/\ell_{\rm loop}$, where the internal loop structure is irrelevant. 
String loops are continuously produced by long-string intercommutation at a rate governed by the VOS model~\eqref{eq:loop_form_rate}. Expressed in terms of the comoving number density, $ \tilde{n}_{\rm loop} \equiv
 a (\eta)^3 n_{\rm loop}$, the loop formation rate is 
\begin{equation}
  \frac{d \tilde{n}_{\rm loop}}{d\eta_i }  = \frac{g \tilde{c} v }{ \xi^3 \alpha}  \frac{1}{ \eta_i^4}  \, . 
\end{equation}

After formation, each global string loop shrinks primarily by radiating Nambu-Goldstone bosons. 
The comoving loop size at conformal time $\eta$ for a loop formed at an earlier time $\eta_i$ evolves as
\begin{equation}
  \ell_{\rm loop}(\eta;\eta_i)
  \;=\;
  \alpha\,\eta_i
    \;-\;\frac{\Gamma_a}{2\pi\,\mathcal{L}_i}\,
    (\eta-\eta_i) \ , 
  \label{eq:ell_loop}
\end{equation}
where the logarithmic factor is defined as $\mathcal{L}_i \equiv \ln (a_i \xi \eta_i / \delta)$.
We treat this logarithm as approximately constant and use the correlation length as the characteristic scale instead of the loop radius, since they are of the same order. 
The parameter $\Gamma_a \simeq 65$~\cite{Battye:1997jk,Vilenkin:2000jqa} is the dimensionless Nambu-Goldstone radiation efficiency\,\footnote{For Nambu-Goto strings, the string loop shrinks predominantly due to gravitational wave radiation, and one can replace $\Gamma_a/(2\pi\mathcal{L}_i)$ with the gravitational radiation power parameter $\Gamma G\mu$.}, which is defined by the energy loss rate $dE/dt = -\Gamma_a f_a^2 / 2$.

\subsection{Unequal Time Correlators}
\label{sec:uetc}
Cosmological observables sourced by cosmic strings are computed from
the unequal-time correlators (UETCs) of the string stress-energy
tensor in Fourier space~\cite{Turok:1996ud,Turok:1996wa,Pen:1997ae}. These correlators encode the statistical properties of the string network and enter as source terms in the evolution equations for metric and matter
perturbations. The UETC is obtained by ensemble averaging over all possible string configurations. In the USM, the network consists of string segments, each described by the single-segment stress-energy tensor $T^{(1)}_{\mu\nu}({\bf k},\eta)$ given in \cref{eq:T00,eq:T_ij}. 
The ensemble average is realized by averaging this single-segment tensor over all possible configurations of the segment, 
namely the position $\mathbf{x}_0$ of its midpoint, its orientation $\hat{\mathbf{x}}$, and 
its velocity direction $\hat{\mathbf{v}}$~\cite{Pen:1997ae,Pogosian:1999np}:
\begin{equation}
\label{eq:uetcdef}
    \langle T_{\mu \nu}({\bf k},\eta_1) T_{\rho \sigma}^{*}({\bf k}',\eta_2)\rangle \; = \; 
 \frac{N}{V_\ell}\int d^3{\bf x}_0\, \int \frac{d^2\hat{\bf x}}{4\pi}\,\int_0^{2\pi} \frac{d\psi}{2\pi} \, T^{(1)}_{\mu \nu}({\bf k},\eta_1) T_{\rho \sigma}^{*(1)}({\bf k}',\eta_2) \, ,
\end{equation}
where $V_{\ell}=\int d^3{\bf x}_0$ is the comoving volume over which the segment position is distributed, $N$ is the number of string segments in this volume, and $\psi$ parametrizes the velocity direction $\hat{\mathbf{v}}$ in the plane transverse to the segment (see \cref{eq:B3}). The superscript $(1)$ emphasizes that $T^{(1)}_{\mu\nu}$ describes a single segment. Since different segments sit at uncorrelated positions, their cross-correlations vanish upon averaging and the network correlator is the incoherent sum of $N$ single-segment contributions, which is the origin of the prefactor $N/V_\ell$. In the rest of the paper we drop the superscript for simplicity of notation.
The segment position enters $T^{(1)}_{\mu\nu}({\bf k},\eta)$ only through the phase factor $e^{-i{\bf k}\cdot{\bf x}_0}$ in \cref{eq:T00}, so the $\mathbf{x}_0$ integration enforces statistical homogeneity and produces a momentum-conserving delta function,
\begin{equation}
  \bigl\langle T_{\mu\nu}^{}(\mathbf{k},\eta_1)\,
    T_{\rho\sigma}^{*}(\mathbf{k}',\eta_2)\bigr\rangle
  \;=\;
  (2\pi)^{3}\,\delta^{(3)}(\mathbf{k}-\mathbf{k}')\;
  T_{\mu\nu,\rho\sigma}(k,\eta_1,\eta_2)\,,
  \label{eq:UETC_delta}
\end{equation}
which defines the UETC tensor $T_{\mu\nu,\rho\sigma}(k,\eta_1,\eta_2)$. 

In the long-wavelength regime, long strings and loops
are spatially uncorrelated, and the total UETC of the network is
simply the sum of the two contributions:
\begin{equation}
  T_{\mu\nu,\rho\sigma}^{\rm tot}(k,\eta_1,\eta_2)
  \;=\;
  T_{\mu\nu,\rho\sigma}^{\rm long}(k,\eta_1,\eta_2)
  \;+\;
  T_{\mu\nu,\rho\sigma}^{\rm loop}(k,\eta_1,\eta_2)\,.
  \label{eq:UETC_total}
\end{equation}
This decomposition lets us compute the two contributions to the
gravitational wave spectrum separately. The two contributions differ only in the number of segments per comoving volume, $N/V_\ell$, entering \cref{eq:uetcdef}.

For long strings, the scaling property implies one segment per correlation volume, so $N=1$ with $V_{\ell}=\max{\left[(\xi\eta_1)^3,(\xi\eta_2)^3\right]}$, the larger of the two correlation volumes at $\eta_1$ and $\eta_2$~\cite{Avgoustidis:2012gb}.

For loops, the formation rate in \cref{eq:loop_form_rate} indicates that multiple loops are produced at a given time. We therefore consider the loops formed within a time interval $[\eta_i,\eta_i+\Delta\eta_i]$, whose number per comoving volume is $N/V_\ell=(d\tilde{n}_{\rm loop}/d\eta_i)\,\Delta\eta_i$, so that
\begin{equation}
     \langle T_{\mu \nu}({\bf k},\eta_1) T_{\rho \sigma}^*({\bf k}',\eta_2)\rangle^{\rm loop} \; = \;\;\frac{d \tilde{n}_{\rm loop}}{d\eta_i}\Delta \eta_i\int d^3{\bf x}_0\, \int \frac{d^2\hat{\bf x}}{4\pi}\,\int_0^{2\pi} \frac{d\psi}{2\pi} \,  T_{\mu \nu}({\bf k},\eta_1) T_{\rho \sigma}^{*}({\bf k}',\eta_2) \, ,
\end{equation}
where $\eta_1,\eta_2\in[\eta_i,\eta_i+\Delta\eta_i]$. The interval $\Delta\eta_i$ will be taken to be infinitesimal when we integrate over the loop production history in \cref{subsec:GW_loop}.

The full USM expressions for the UETCs are given in \cref{app:UETC}. In the infrared regime $k\,\ell \lesssim 1$ relevant for our analytic
estimates, the UETCs factorize into a kinematic tensor times a scalar
function:
\begin{equation}
  T_{\mu\nu,\rho\sigma}(k,\eta_1,\eta_2)
  \;=\;
  F_{\mu\nu,\rho\sigma}(v)\;\cdot\;
  T(k,\eta_1,\eta_2)\,,
  \label{eq:UETC_factorised}
\end{equation}
where $F_{\mu\nu,\rho\sigma}(v)$ encodes the Lorentz boost factor and the angular averages over
segment orientations and velocity directions (explicit expressions in
\cref{eq:B10}). The only component relevant for tensor modes is $F_{12,12}$, which takes the simple form
\begin{equation}
  F_{12,12} = \frac{v^4-v^2+1}{15(1-v^2)}. 
  \label{eq:F1212}
\end{equation}
The scalar UETC takes different forms for the two components of the network: for \textit{long strings} we have,
\begin{equation}
 T^{\rm long}(k,\eta_1,\eta_2)
  \;=\;
  \mu_{\rm eff}(\eta_1) \cdot  \mu_{\rm eff}(\eta_2) \cdot \frac{\ell_{\rm long} (\eta_1) \cdot \ell_{\rm long} (\eta_2)}{ (\max(\ell_{\rm long} (\eta_1),\ell_{\rm long} (\eta_2)))^3},
  \qquad  k\ell_{\rm long} \lesssim 1\,,
  \label{eq:T_long}
\end{equation}
where $\ell_{\rm long}(\eta)\equiv\xi\eta$.

For \textit{loops}, the scalar UETC is
\begin{equation}
  T^{\rm loop}(k,\eta_1,\eta_2)
  \;=\;
  \frac{d\tilde{n}_{\rm loop}}{d\eta_i} \Delta \eta_i \cdot \mu_{\rm eff}(\eta_1) \cdot \mu_{\rm eff}(\eta_2) \cdot \ell_{\rm loop}(\eta_1)\cdot \ell_{\rm loop}(\eta_2),
  \qquad k\,\ell_{\rm loop} \lesssim 1\,,
  \label{eq:T_loop}
\end{equation}
where $\eta_1,\eta_2\in[\eta_i,\eta_i+\Delta \eta_i]$. The featureless-segment approximation underlying these expressions breaks down at sub-correlation-length scales, where kinks, cusps, and small-scale wiggles become important. We address this ultraviolet regime in \cref{sec:uvregime}.

\section{Gravitational Waves from Cosmic Strings}
\label{sec:gwstrings}

In this section, we compute the stochastic GW background sourced by a
scaling cosmic string network. GWs are treated as transverse-traceless
tensor perturbations on an FLRW background, sourced by the anisotropic
stress of the network and propagated with retarded Green's functions.
Combining these Green's functions with the UETCs of \cref{sec:cs}, we
obtain a master formula for the present-day GW spectral density
$\Omega_{\rm GW,0}(k)$, which organizes the spectrum into three
regimes: superhorizon~(I), subhorizon super-correlation~(II), and
sub-correlation~(III). In the infrared ($k\lesssim\ell_c^{-1}$,
regimes~I and~II), the spectrum is model-independent and computed
analytically within the USM for both long strings and loops, whose
contributions turn out to be comparable. In the
ultraviolet ($k\gtrsim\ell_c^{-1}$, regime~III), the straight-segment
approximation breaks down, and we develop a data-driven method that
maps the instantaneous radiation power spectrum measured in lattice
simulations onto the same UETC formalism.

\subsection{Green's Functions for Tensor Modes}
\label{sec:gftensormodes}
We now begin the computation of the stochastic GW background sourced by a cosmic-string network in linear perturbation theory, treating GWs as tensor perturbations on a spatially flat FLRW background. The perturbed
spatially flat FLRW metric in conformal time~$\eta$ reads
\begin{equation} 
\mathrm{d}s^2 \; = \; a^2(\eta) \left[ \mathrm{d}\eta^2 - (\delta_{ij} + h_{ij}) \, \mathrm{d}x^i \mathrm{d}x^j \right]\, , 
\label{eq:c1} 
\end{equation} 
where $h_{ij}$ is the tensor metric perturbation in the
transverse-traceless (TT) gauge,
$h^{i}{}_{i}=\partial_{i}h^{i}{}_{j}=0$. At linear order, $h_{ij}$
satisfies the wave equation~\cite{Baumann:2022mni}
\begin{equation}
  h''_{ij}+2\mathcal{H}\,h'_{ij}-\nabla^{2}h_{ij}
  \;=\;16\pi G\,a^{2}\,\Pi_{ij}^{\rm TT}\,,
  \label{eq:hij_eom}
\end{equation}
where $\mathcal{H}\equiv a'/a$ is the conformal Hubble parameter,
primes denote $\partial_{\eta}$, and
$\Pi_{ij}^{\rm TT}$ is the transverse-traceless projection of the
anisotropic stress sourced by the string network. The relation between
$\Pi_{ij}^{\rm TT}$ and the string stress-energy tensor~$T_{ij}$ is
given by the TT projector,
\begin{equation}
  \Pi_{ij}^{\rm TT}(\mathbf{k},\eta)
  \;=\;a^{-2}\Lambda_{ij,lm}(\hat{\mathbf{k}})\,
  T_{lm}(\mathbf{k},\eta)\,,
  \label{eq:TTprojection}
\end{equation}
where
$\Lambda_{ij,lm}=P_{il}P_{jm}-\tfrac{1}{2}P_{ij}P_{lm}$ with
$P_{ij}=\delta_{ij}-\hat{k}_{i}\hat{k}_{j}$.

To solve~\eqref{eq:hij_eom}, we decompose $h_{ij}$ into Fourier modes and
introduce the dimensionless variable $u\equiv k\eta$. The retarded
solution is
\begin{equation}
  h_{ij}(\mathbf{k},\eta)
  \;=\;\frac{16\pi G}{k^{2}}
  \int_{u_i}^{u}\!dv\;
  \mathcal{G}(u,v)\;a^{2}(v/k)\;
  \Pi_{ij}^{\rm TT}(\mathbf{k},v/k)\,,
  \label{eq:greensol}
\end{equation}
where $u_{i}$ corresponds to the onset of GW production and
$\mathcal{G}(u,v)$ is the retarded Green's function satisfying
\begin{equation}
  \mathcal{G}''(u,v)
  +\frac{2\mathcal{H}}{k}\,\mathcal{G}'(u,v)
  +\mathcal{G}(u,v)
  \;=\;\delta(u-v)\,,
  \qquad
  \mathcal{G}(u,v)=0\;\text{for}\;u<v\,,
  \label{eq:Greq}
\end{equation}
with primes now denoting $\partial_{u}$. The analytic solutions for
$\mathcal{G}$ in the radiation- and matter-dominated eras are
shown here,
\begin{equation}
        \mathcal{G}(u, v) \; = \; \Theta(u - v) \cdot \frac{v \sin(u-v)}{u} ~(\rm{RD})\, , 
\label{eq:greenRD}\end{equation}
\begin{equation}
    \mathcal{G}(u, v) \; = \; \Theta(u - v) \cdot \left[\frac{v(v-u)\cos(u-v) + v(1+uv)\sin(u-v)}{u^3} \right]~(\rm{MD}) \, .
\label{eq:greenMD}\end{equation}
The detailed derivation is given in Appendix~\ref{app:greenfunctions}.

\subsection{Gravitational Wave Spectral Density}
\label{sec:gwspectraldensity}
The two radiative degrees of freedom are identified by decomposing the
tensor perturbation in the polarisation basis,
\begin{equation}
  h_{ij}(\mathbf{k},\eta)
  \;=\;\sum_{\lambda=+,\times}
  h_{\lambda}(\mathbf{k},\eta)\;
  \varepsilon_{ij}^{\lambda}(\hat{\mathbf{k}})\,,
  \label{eq:h_pol}
\end{equation}
where the polarisation tensors satisfy
$\varepsilon_{ij}^{\lambda}\,k^{j}=0$,
$\delta^{ij}\varepsilon_{ij}^{\lambda}=0$, and
$\varepsilon_{ij}^{\lambda}\,
\varepsilon^{\lambda'\,ij}=\delta^{\lambda\lambda'}$.
Combining the $a^{-2}$ factor in the TT
projection~\eqref{eq:TTprojection} with the $a^{2}$ prefactor in the
Green's function solution~\eqref{eq:greensol}, the scale-factor
dependence cancels and the retarded solution becomes
\begin{equation}
  h_{ij}(\mathbf{k}, \eta)
  \;=\;\frac{16\pi G}{k^{2}}
  \int_{u_i}^{u}\!dv\;
  \mathcal{G}(u,v)\;
  \Lambda_{ij,lm}(\hat{\mathbf{k}})\,
  T_{lm}\!\left(\mathbf{k},\tfrac{v}{k}\right),
  \label{eq:h_solution}
\end{equation}
or equivalently 
\begin{equation}
  h_{\lambda}(\mathbf{k},\eta)
  \;=\;\frac{16\pi G}{k^{2}}
  \int_{u_i}^{u}\!dv\;
  \mathcal{G}(u,v)\;
  \Lambda_{ij,lm}(\hat{\mathbf{k}})\,
 \e^\lambda_{ij}\, T_{lm}\!\left(\mathbf{k},\tfrac{v}{k}\right).
  \label{eq:h_solution_pol}
\end{equation}

The present-day fractional GW energy density per logarithmic
wavenumber is
\begin{equation}
  \Omega_{\rm GW,0}(k)
  \;=\;\frac{1}{12}\left(\frac{k}{a_0 H_0}\right)^{\!2}
  \frac{k^{3}}{2\pi^{2}}\,
  \sum_{\lambda}
  \bigl\langle|h_{\lambda}(\mathbf{k},\eta_0)|^{2}\bigr\rangle\,,
  \label{eq:OmGW_def}
\end{equation}
where $H_0$ is the present-day Hubble parameter. Substituting
\cref{eq:h_solution_pol} into~\eqref{eq:OmGW_def} and expressing the
ensemble average in terms of the string
UETC~\eqref{eq:UETC_delta}, the polarization sum contracts with the
kinematic tensor of the UETC to
give $\sum_{\lambda} \varepsilon_{ij}^{\lambda} \varepsilon_{lm}^{\lambda\,*}F^{ij,lm}(v) =4\,F_{12,12}(v)$.
Combining all factors, we arrive at the master formula for the GW
spectrum:
\begin{equation}
\begin{aligned}
  \Omega_{\rm GW,0}(k)
  &\;=\;\frac{128\,G^{2}\,k}{3\,a_0^{2}\,H_0^{2}}\;
  F_{12,12}(v)
  \int_{u_i}^{u_{f}}\!dw'\!
  \int_{u_i}^{u_{f}}\!dw''\;
  \mathcal{G}(u_0,w')\,\mathcal{G}(u_0,w'')\;
  T\!\left(k,\frac{w'}{k},\frac{w''}{k}\right)\\ 
\end{aligned}
  \label{eq:OmGW_master}
\end{equation}
where $u_0 = k\eta_0$, and
$T = T^{\rm long} + T^{\rm loop}$ is the total scalar UETC defined in
\cref{eq:T_long,eq:T_loop}, and $F_{12,12}(v)$ is given in \cref{eq:F1212} . The integration runs from
$u_i\equiv k\,\eta_{i}$, the onset of the scaling regime,
to $u_{f}\equiv k\,\eta_{f}$, the epoch at which the
string network collapses.

In principle, \cref{eq:OmGW_master} allows us to compute the GW spectrum for an arbitrary source history, but no closed-form Green's 
function exists that spans the full RD-to-MD transition. 
The calculation simplifies in some limiting cases. For modes that enter the horizon well before matter--radiation equality, 
the source is effectively active only during the RD era and the RD Green's function \cref{eq:greenRD} applies throughout the integration; 
the GW energy density is evaluated at $\eta_\mathrm{eq}$ and subsequently redshifted as radiation to the present. 
Conversely, for modes that are produced during 
the MD period, the MD Green's function \cref{eq:greenMD} applies throughout the source epoch. For intermediate modes, or when the source spans 
the transition, we integrate $h$ using the RD Green's function up to $\eta_\mathrm{eq}$, match $h$ and $\dot{h}$ to the homogeneous 
free-wave solutions at equality, and then propagate it forward using those solutions.

The spectrum is organized by two dimensionless ratios: the
horizon-entry parameter~$k\eta_f$ and the correlation-entry
parameter~$k\,\ell ( \eta_f)$ at $\eta_f$, divided into three regimes:
\begin{align}
  &\text{(I)\;superhorizon:}
    &&k\eta_f \ll 1,\quad k\,\ell(\eta_f) \ll 1\,,
    \label{eq:regime_I}\\[4pt]
  &\text{(II)\;subhorizon, super-correlation:}
    &&k\eta_f \gg 1,\quad k\,\ell(\eta_f) \ll 1\,,
    \label{eq:regime_II}\\[4pt]
  &\text{(III)\;sub-correlation:}
    &&k\eta_f \gg 1,\quad k\,\ell(\eta_f) \gg 1\,.
    \label{eq:regime_III}
\end{align}
Note that we use the string terminated time $\eta_f$ as a reference point to classify the regimes, since the gravitational wave contribution is dominant at that time. 
In regime~(I) the mode is outside the horizon at $\eta_f$. In regime~(II), the mode is inside the horizon but does not resolve the internal structure of the network: the string segments appear as featureless sources with stochastic orientations and velocities. These are the regimes in which the straight-segment (USM) description is controlled, and where the model-independent IR spectrum is determined. In regime~(III) the mode probes scales below the correlation length, where kinks, cusps, and small-scale wiggles become relevant and the featureless-segment approximation breaks down. But there still exist the IR contributions produced at earlier time, $\eta \ll \eta_f$, in regime (III), and it is subdominant compared to the UV spectra. We address this ultraviolet regime with data-driven input in Section~\ref{sec:uvregime}.

\subsection{Gravitational Waves from Long Strings}
\label{subsec:longstrings}
In this subsection, we derive the GW spectrum sourced by long global
strings in the scaling regime, evaluating the master
formula~\eqref{eq:OmGW_master} in the three regimes (I)--(III)
classified above. The two physical inputs are the logarithmically
running effective tension,
$\mu_{\rm eff}(\eta)=\pi f_a^{2}\ln(\ell_c a(\eta)/\delta)$ of
\cref{eq:mu_eff}, and the network decay time $\eta_f$, which we
specify below. In regimes (I) and (II) the result is
model-independent, fully fixed by the network parameters $(\xi, v)$.
The analogous results for gauge strings follow by replacing
$\mu_{\rm eff}\to\mu$ and are presented in \cref{app:gaugestrings},
where they are validated against Nambu-Goto simulations.

We emphasize that the calculation in this subsection is valid only in
the IR region, $k\ell_c\lesssim 1$. As shown in \cref{eq:T00}, the IR
cutoff of the logarithm in $\mu_{\rm eff}$ is set by the smaller of
the physical wavelength and the physical correlation length,
$\min(a/k,\,a\ell_c)$. In the long-wavelength limit relevant for
regimes~(I) and~(II), the correlation length sets the cutoff and the
tension reduces to the $k$-independent form of \cref{eq:mu_eff}. For
$k\ell_c\gtrsim 1$, the wavelength itself sets the cutoff, the
tension becomes explicitly $k$-dependent, and sub-correlation-length
structure dominates the source; in regime~(III) we therefore compute
only the IR contribution, deferring the full UV spectrum to the
data-driven treatment of \cref{sec:uvregime}.

For global (axion) strings, the network terminates when the axion
mass becomes cosmologically relevant at $H(\eta_f)\simeq m_a$, at
which point domain-wall dynamics becomes essential. If the resulting
string--wall system is unstable, the wall tension pulls the strings
together, causing the network to fragment and decay via axion
radiation. We model this decay as instantaneous at~$\eta_f$ and focus
on the GW signal from the scaling network prior to the network
collapse. Since~$\eta_f$ is set by the axion mass, it is a free
parameter that controls the peak frequency of the GW spectrum, as we
discuss in \cref{sec:global_spectrum}.

In regime (I), modes are superhorizon at the network decay time,
$k\eta_f\ll 1$. At these scales, the scaling network sources a
white-noise metric perturbation, $h(k,\eta)\sim G\mu_{\rm eff}$,
independent of~$k$. Depending on the decay time and the wavelength,
a mode either re-enters the horizon during RD ($\eta_f<\eta_{\rm eq}$,
$k\eta_{\rm eq}>1$) or during MD ($k\eta_{\rm eq}<1<k\eta_0$), or
remains superhorizon today ($k\eta_0<1$). Upon re-entry at
$\eta_{\rm re}\sim k^{-1}$, the mode oscillates with the transfer
function $h(k,\eta)\simeq h(k,\eta_{\rm re})\sin(k\eta)/(k\eta)$ in
the radiation-dominated background, or
$h(k,\eta)\simeq -3\,h(k,\eta_{\rm re})\cos(k\eta)/(k\eta)^{2}$ in the
matter-dominated background~\cite{Wu:1998mr,Gouttenoire:2019kij},
while a mode that has not re-entered remains frozen at its
superhorizon value. From~\eqref{eq:OmGW_def} we obtain the following:
\begin{equation}
  \Omega_{\rm GW,0}^{\rm(I)}(k)
  \;=\;\frac{128\,(G\mu_{\rm eff}(\eta_f))^{2}}{9 \,{a_0^{2}H_0^{2} \eta_0^{2}}}\;\xi^{-1}\;
  F_{12,12}(v)\;
  \;\times \begin{cases} \frac{1}{3}
  (k\eta_f)^3\,,
  \qquad 1 < k\eta_{\rm eq}\,,\ \eta_f < \eta_{\rm eq} \,,\\
  \\
  \frac{3}{2} \left(\frac{\eta_f}{\eta_0}\right)^2 \cdot k\eta_f \,, 
  \qquad k\eta_{\rm eq} < 1 < k\eta_0\,,\ \eta_f < \eta_{\rm eq}\,,\\
  \\
  \frac{1}{6} \left(\frac{\eta_0}{\eta_f}\right)^{2} \cdot (k\eta_f)^5 \,, 
  \qquad  k\eta_0 < 1\,,\ \eta_{\rm eq} < \eta_f\,,\\
  \\
  \frac{1}{12} \left(\frac{\eta_f}{\eta_0}\right)^{2} \cdot k\eta_f \,, 
  \qquad  k\eta_f < 1< k\eta_0\,,\ \eta_{\rm eq} < \eta_f
  \end{cases}
  \label{eq:Omega_gw_I}
\end{equation}
where $\mu_{\rm eff}(\eta_f)=\pi f_a^{2} \ln(a_f\xi\eta_f/\delta_c)$.

In regime (II) (subhorizon, super-correlation),
$1<k\eta_f<\xi^{-1}$, the mode is inside the horizon while the
network is still active, but its wavelength exceeds the correlation
length. Each correlation volume acts as an independent, unresolved
source: the anisotropic stress remains white noise in position
space, and, unlike regime~(I), the mode oscillates and redshifts
while being sourced, yielding the characteristic
$\Omega_{\rm GW}\propto k^{3}$ spectrum of uncorrelated sources\footnote{However, for the case of $\xi\ll\mathcal{O}(0.1)$, part of the MD spectrum in regime (II) can scale as $k$ instead of $k^3$ in the range of $0.1^{-1}/\eta_f\lesssim k<\xi^{-1}/\eta_f$.}.
Evaluating the master formula~\eqref{eq:OmGW_master} gives
\begin{equation}
  \Omega_{\rm GW,0}^{\rm(II)}(k)
  \;=\;\frac{128\,(G\mu_{\rm eff}(\eta_f))^{2}}{3}\;\xi^{-1}\;
  F_{12,12}(v)\;(k\eta_f)^{3}
  \;\times\;
  \begin{dcases}
    \dfrac{\Omega_r}{9}
    \left(\dfrac{g_{*s}(T_f)}{g_{*s,0}}\right)^{\!-1/3},
    & 1 < k\eta_f\,,\ \eta_f<\eta_{\rm eq}\,,\\[6pt]
    \dfrac{\Omega_m^{2}\,a_0^{2}H_0^{2}\,\eta_f^{2}}{32\,}\,,
    & 1 < k\eta_f\,,\ \eta_f>\eta_{\rm eq}\,,
  \end{dcases}
  \label{eq:Omega_gw_II}
\end{equation}
where $g_{*s}(T)$ is the effective number of entropic degrees of
freedom, $T_f$ the temperature at~$\eta_f$, and $\Omega_r$,
$\Omega_m$ the present-day radiation and matter density parameters.

In regime~(III) (sub-correlation), a mode with $k\ell(\eta_f)\gtrsim 1$
was nevertheless super-correlation at earlier times: since the
comoving correlation length grows as $\ell(\eta)=\xi\eta$, the same
mode satisfies $k\ell(\eta)\lesssim 1$ for
$\eta<\eta_{f,*}\equiv(\xi k)^{-1}$. The GWs sourced during this
earlier period constitutes an IR contribution to the sub-correlation
spectrum that is still captured by the USM. We compute it by
truncating the UETC integration at $u_{f,*}=k\eta_{f,*}=\xi^{-1}$
(i.e.,\ setting $T(k,\frac{w'}{k},\frac{w''}{k})=0$ for
$w',w''>\xi^{-1}$). The resulting IR contribution is
model-independent, but the full regime~(III) spectrum can be
dominated by the UV contribution from sub-correlation-length
structure (e.g., small-scale wiggles), which we address in
Section~\ref{sec:uvregime}. For completeness, we present the
analytical result of the IR contribution in regime~(III):
\begin{equation}
\label{eq:Omega_gw_III_IR}
    \Omega_{\rm GW,0}^{\text{(III),IR}}(k) = \frac{128( G \mu_{\rm eff}(\eta_{f,*}))^2}{3}\xi^{-4} F_{12,12}(v)    \times 
    \begin{dcases}
    \dfrac{\Omega_r}{9}\!
\left(\dfrac{g_{*s}(T_f)}{g_{*s,0}}\right)^{-1/3}, & \xi^{-1} < k\eta_f\,,\ \eta_f<\eta_{\rm eq}\, ,\\[2.2ex]
\dfrac{\Omega_m^2a_0^2H_0^2\eta_f^2}{32 }\dfrac{\xi^{-2}}{k^2\eta_f^2}, & k\eta_{\rm eq} < \xi^{-1} < k\eta_f\, ,
\end{dcases}
\end{equation}
where $\mu_{\rm eff}(\eta_{f,*})$ is the effective string tension
in~\cref{eq:mu_eff} evaluated at the characteristic time
$\eta_{f,*}$, at which the mode wavelength equals the correlation
length, and the $k^{-2}$ dependence of the GW intensity in the $\eta_f > \eta_{\rm eq}$ case arises due to the quadratic dependence of the scale factor on the conformal time $\eta$. Comparing \cref{eq:Omega_gw_III_IR} with~\cref{eq:Omega_gw_II},
the IR contribution in regime~(III) has different spectral shapes in
different backgrounds: in RD it is a plateau,
$\Omega_{\rm GW,0}^{\text{(III),IR}}(k)
=\Omega_{\rm GW,0}^{\text{(II)}}(\xi^{-1}/\eta_f)
=\max\!\left(\Omega_{\rm GW,0}^{\text{(II)}}(k)\right)$. The
$k^{0}$ plateau in RD follows from the scaling property of the string
network: the string energy density remains a fixed fraction of the
background, so the GW energy produced per Hubble time is a constant
fraction of the total. For each mode~$k$, the dominant production
occurs around $\eta_{f,*}\propto k^{-1}$, and during RD the emitted
GWs redshift in the same way as the background radiation, so every
mode retains the same fractional energy density today. In MD, in contrast,
$\Omega_{\rm GW,0}^{\text{(III),IR}}(k)/
\Omega_{\rm GW,0}^{\text{(II)}}(\xi^{-1}/\eta_f)
=\xi^{-2}/(k\eta_f)^2$, indicating a power-law suppression
$\propto k^{-2}$.

Finally, we comment on the differences between the GW spectra from
long global and gauge strings. The gauge-string results follow
directly from
\cref{eq:Omega_gw_I,eq:Omega_gw_II,eq:Omega_gw_III_IR} by setting
$\mathcal{L}_f\to 1$, i.e., $\mu_{\rm eff}\to\mu$. In regimes~(I)
and~(II), $\mu_{\rm eff}$ is independent of~$k$ in the
long-wavelength limit, so the global-string spectrum retains the
same frequency dependence as the gauge-string one, with the overall
amplitude enhanced by $\mathcal{L}_f^{2}$. Since
$\mathcal{L}_f=\mathcal{O}(100)$ for a network with
$f_a\sim10^{15}\,$GeV decaying deep in RD, this enhancement reaches
$\sim4$ orders of magnitude. In regime~(III), by contrast, the
effective tension is evaluated at the $k$-dependent time
$\eta_{f,*}=(\xi k)^{-1}$, so the logarithm becomes scale-dependent,
$\mathcal{L}_f\to\mathcal{L}_{f,*}
=\ln(a_{f,*}\,\xi\,\eta_{f,*}/\delta_c)$, which decreases with~$k$.
Explicitly, in RD,
$\mathcal{L}_{f,*}
=\ln\!\left[(g_{*s}/g_{*s,0})^{-1/6}\sqrt{\Omega_r}\,
H_0\,\xi^{-1}\,\delta_c^{-1}/k^2\right]$, while in MD,
$\mathcal{L}_{f,*}
=\ln\!\left[\tfrac{1}{4}\Omega_m H_0^2 \xi^{-2}\,
\delta_c^{-1}/k^3\right]$. As a result, the regime~(III) IR spectrum
of global strings acquires a mild logarithmic tilt: the RD plateau
is not exactly flat but decreases logarithmically with~$k$, in
contrast to the scale-invariant plateau of gauge strings.

\subsection{Gravitational Waves from Loops}
\label{subsec:GW_loop}

Cosmic string loops radiate GWs. In this subsection, we derive the
infrared GW spectrum ($k\ll 1/\ell_{\rm loop}$) from global string
loops, using the extension of the USM to shrinking loop segments
developed in \cref{sec:USLM}. Since the lifetime of a global string
loop is shorter than the Hubble timescale, loops are continuously
produced and decay, and, as we show below, they source a universal
white-noise IR spectrum. The ultraviolet spectrum
($k\gtrsim 1/\ell_{\rm loop}$), dominated by loop substructure such
as cusps, kinks, and kink-kink collisions, is widely studied in the
literature, e.g., using the VOS model and the harmonic
method~\cite{Cui:2018rwi}. The IR part, on the other hand, lacks
analytical understanding and is difficult to extract from lattice
simulations: modes near the simulation box scale are sampled by only
a few independent wavelengths and thus suffer from large
uncertainties.

The loop GW spectrum is computed with the same Green's function
formalism as for long strings. The key difference is that loops are
continuously produced by long-string intercommutation and decay
within a Hubble time: at any given time, a new generation of loops
is created and radiates GWs during its short lifetime. In the long
wavelength limit, $k\ell_{\rm loop}\ll1$, each loop generation is
modeled by shrinking segments with the UETC given in
\cref{eq:T_loop}, following the same two-point statistics as
long-string segments,
$T^{\rm loop}_{12,12}(k,\eta,\eta')=T^{\rm loop}(k,\eta,\eta')\,
F_{12,12}(v)$, with the loop size evolution $\ell_{\rm loop}(\eta)$
of \cref{eq:ell_loop}. Substituting the UETC into the master
formula~\eqref{eq:OmGW_master} and summing over the loop
generations, the total GW spectrum reads
\begin{equation}
\begin{aligned}
    \Omega_{\rm GW,0}(k) \;=\; &\frac{128\,G^2}{3\,a_0^2 H_0^2}\,
    F_{12,12}(v)
    \int_{u_{\rm ini}}^{u_f}\! du_i\,
    \frac{d\tilde{n}_{\rm loop}}{d\eta_i}
    \int_{u_i}^{u_d}\!dw'\!
    \int_{u_i}^{u_d}\!dw''\;
    \mathcal{G}(u_0,w')\,\mathcal{G}(u_0,w'')\\[2pt]
    &\times
    \mu_{\rm eff}\!\left(\tfrac{w'}{k}\right)
    \mu_{\rm eff}\!\left(\tfrac{w''}{k}\right)
    \ell_{\rm loop}\!\left(\tfrac{w'}{k}\right)
    \ell_{\rm loop}\!\left(\tfrac{w''}{k}\right) \, .
\end{aligned}
\label{eq:OmGW_loop_master}
\end{equation}

The outer integral over
the formation time $u_i=k\eta_i$ captures the continuous GW emission
from the newly generated loops, running from the initial time
$\eta_{\rm ini}$, e.g., when the string network enters the scaling
regime, or the radiation-matter equality $\eta_{\rm eq}$ (if
evaluating the MD-only contribution), to the collapse time of the
network $\eta_f$. The inner double time integral encodes the GW
emission of a single loop generation over its lifetime, with the
upper limit set by the loop decay time
$u_{d}\equiv k\eta_{d}=(1+\frac{2\pi\alpha \mathcal{L}_i}{\Gamma_a})u_i$.\footnote{For the MD-only contribution, $u_d=\text{min}\left((1+\frac{2\pi\alpha \mathcal{L}_i}{\Gamma_a})u_i,u_0\right)$ where $u_0=k\eta_0$, which enforces that the emission ends by the present epoch. }

To evaluate \cref{eq:OmGW_loop_master}, we follow the same procedure
as in \cref{subsec:longstrings}: depending on the regime, the
Green's functions are simplified by Taylor expansion for
superhorizon modes or averaged over one oscillation period for
subhorizon modes, and each mode is then propagated to today with the
appropriate transfer function, according to whether it re-enters the
horizon during RD or MD. Performing the integration over the
formation time, we find that the GW spectra in regimes~(I) and~(II)
are

\begin{equation}
\begin{aligned}
  \Omega_{\rm GW,0}^{\rm(I)}(k)
  \;=\;&\frac{128\,\tilde{c}vg\alpha}{9\,\xi^{3}}\,
  (G\mu_{\rm eff}(\eta_f))^{2}\,
  \frac{F_{12,12}(v)}{(a_0 H_0\,\eta_0)^{2}}
  \left(\frac{\pi \alpha \mathcal{L}_f}{\Gamma_a}\right)^{2}
  \left(1 + \frac{2\pi \alpha \mathcal{L}_f}{3\Gamma_a}\right)^{2}\\[2pt]
  &\times \begin{dcases}
    \frac{9}{2} \left(\frac{\eta_f}{\eta_0} \right)^{2} k\eta_f\,,
    &  k\eta_f  < 1< k\eta_{0}\,,\\[1.2ex]
    \frac{1}{2} (k\eta_f)^{3}\,,
    &  k\eta_f <1< k\eta_{\rm eq}\,,\\[1.2ex]
    \frac{1}{36} \left(\frac{\eta_0}{\eta_f}\right)^{2} (k\eta_f)^{5}\,,
    &  k\eta_f < k\eta_{0} < 1\,,\\ 
    \frac{1}{8} \left( \frac{\eta_f}{\eta_0} \right)^2 k\eta_f , & k\eta_f < 1 < k\eta_0 \,,\, \eta_{\rm eq} < \eta_f
  \end{dcases}
\end{aligned}
\label{eq:Omegagw_loop_I}
\end{equation}
\begin{equation}
\begin{aligned}
  \Omega_{\rm GW,0}^{\rm(II)}(k)
  \;=\;&\frac{64\,\tilde{c}vg\alpha}{9\,\xi^{3}}\,
  (G\mu_{\rm eff}(\eta_f))^{2}\,
  F_{12,12}(v)
  \left(\frac{\pi \alpha \mathcal{L}_f}{\Gamma_a}\right)^{2}
  (k\eta_f)^{3}\\[2pt]
  &\times \begin{dcases}
    \Omega_r \left(\frac{g_{*s}(T_f)}{g_{*s,0}}\right)^{\!-1/3}
    \left(1 + \frac{2\pi \alpha \mathcal{L}_f}{3\Gamma_a}\right)^{2},
    & 1 < k\eta_f\,,\ \eta_f<\eta_{\rm eq}\,,\\[1.2ex]
    \frac{\Omega_m^{2}\,a_0^{2}H_0^{2}\,\eta_f^{2}}{8}
    \left(1+\frac{4\pi \alpha \mathcal{L}_f}{3\Gamma_a}
    +\frac{2\pi^{2} \alpha^{2} \mathcal{L}_f^{2}}{3\Gamma_a^{2}}\right)^{2},
    & 1 < k\eta_f\,,\ \eta_f>\eta_{\rm eq}\,,
  \end{dcases}
\end{aligned}
\end{equation} 
\begin{equation} 
\begin{aligned}
  \Omega_{\rm GW,0}^{\rm(III)}(k)
  \;=\;&\frac{64\,\tilde{c}vg\alpha}{9\,\xi^{6}}\,
  (G\mu_{\rm eff}(\eta_{f,\ast}))^{2}\,
  F_{12,12}(v)
  \left(\frac{\pi \alpha \mathcal{L}_{f,\ast}}{\Gamma_a}\right)^{2}
  \\[2pt]
  &\times \begin{dcases}
    \Omega_r \left(\frac{g_{*s}(T_f)}{g_{*s,0}}\right)^{\!-1/3}
    \left(1 + \frac{2\pi \alpha \mathcal{L}_{f,\ast}}{3\Gamma_a}\right)^{2},
    & \xi^{-1} < k\eta_f\,,\ \eta_f<\eta_{\rm eq}\,,\\[1.2ex]
    \frac{\Omega_m^{2}\,a_0^{2}H_0^{2}\,\eta_f^{2}}{8}\dfrac{\xi^{-2}}{k^2\eta_f^2}
    \left(1+\frac{4\pi \alpha \mathcal{L}_{f,\ast}}{3\Gamma_a}
    +\frac{2\pi^{2} \alpha^{2} \mathcal{L}_{f,\ast}^{2}}{3\Gamma_a^{2}}\right)^{2},
    &  k\eta_{\rm eq} < \xi^{-1} < k\eta_f\,.\,
  \end{dcases}
\end{aligned}
\label{eq:Omegagw_loop}
\end{equation}

Compared to the long-string contribution in regime~(II),
\cref{eq:Omega_gw_II}, the loop amplitude formally scales as
$\mathcal{L}_f^{4}$: two powers of $\mathcal{L}_f$ come from the
effective string tension $\mu_{\rm eff}^{2}$, as in the long-string
case, while the factor $(\pi\alpha\mathcal{L}_f/\Gamma_a)^{2}$
carries two additional powers, originating from the loop lifetime
set by the Nambu-Goldstone radiation shrinkage rate
in~\cref{eq:ell_loop}. We emphasize, however, that despite the
$\mathcal{L}_f^{4}$ scaling, the factor
$\pi\alpha\mathcal{L}_f/\Gamma_a=\mathcal{O}(1)$ for realistic
parameters ($\alpha\simeq0.1$, $\Gamma_a\simeq65$, and
$\mathcal{L}_f=\mathcal{O}(100)$), so the loop contribution is of
the same order as the long-string contribution. This is confirmed
in \cref{sec:global_spectrum}, where the two contributions are found
to be roughly equal for the parameters we choose. The spectral shape remains $\propto k^{3}$ for both components: in
the long-wavelength limit, long strings and loops alike behave as
spatially uncorrelated sources, whose anisotropic stress is white
noise in position space. What differs is the temporal decorrelation
of the source, which generates the unequal-time structure of the
UETC and hence the GW emission: for long strings it is governed by
the random motion of segments and the growth of the correlation
length on the timescale $\sim\xi\eta$ (\cref{sec:USLM}), whereas for
global string loops it is governed by their continuous production
and decay within a Hubble time.

\subsection{Data-driven method for UV spectra}
\label{sec:uvregime}
In the short-wavelength regime, $k \gtrsim 1/\ell$ (regime III), the
unconnected segment and loop model (USLM) approximation for the two-point statistics of long strings and
loops breaks down: the substructure of the network, such as kinks,
cusps, and small-scale wiggles, dominates the source and must be
taken into account. The UV GW spectrum can then be obtained either
by modeling the substructure or using the
instantaneous radiation power spectrum measured in numerical
simulations as input. In this subsection, we develop the latter,
\textit{data-driven} approach: we show how the instantaneous
radiation power spectrum, combined with a transfer function derived
from the UETC formalism, determines the UV GW spectrum.

\paragraph{Theory: the GW two-point function from the source kernel.}
Our goal is to map the instantaneous radiation power spectrum onto the UETC approach to obtain the final GW spectra. We use a general formalism for the UETC below, while \cref{eq:UETC_factorised} only applies in the string segment model.
Starting from the general UETC formalism, the two-point function of the metric perturbation is sourced by the string,
\begin{equation}
\begin{aligned}
    \sum_{\lambda} | h_\lambda(k, \eta)|^2 \; = \; \left(\frac{16\pi G}{k^2}\right)^2\int_{u_i}^{u_f} dw' \int_{u_i}^{u_f} dw'' \, \mathcal{G}(u, w')  \, \mathcal{G}(u, w'')  \sum_{\lambda}\varepsilon_{\lambda}^{ij} \varepsilon_{\lambda}^{*mn} T_{ij,mn}\left(k, \frac{w'}{k}, \frac{w''}{k}\right) \, .
\end{aligned}    
\label{eq:h2}
\end{equation}
The double time integral can be reduced to a simple time integral by the change of variables $\bar{w}=(w'+w'')/2,\,w_-=w'-w''$. The final expression is
\begin{align}
      \sum_{\lambda}|h_{\lambda}(k, \eta)|^{2}
  \;&=\;\left(\frac{16\pi G}{k^{2}}\right)^{\!2}
  \frac{1}{u^{2}}
  \int_{u_i}^{u_f}\!d\bar{w}\;\bar{w}^{2}\,K(\bar{w})\,,~~~~\text{ RD}
  \label{eq:GW_Kernel}\\
  \sum_{\lambda}|h_{\lambda}(k, \eta)|^{2}
  \;&=\;\left(\frac{16\pi G}{k^{2}}\right)^{\!2}
  \frac{1}{u^{4}}
  \int_{u_i}^{u_f}\!d\bar{w}\;\bar{w}^{4}\,K(\bar{w})\,,~~~~\text{ MD}
  \label{eq:GW_Kernel_MD}
\end{align}
where the {\it transfer functions} $\bar{w}^{2}/u^{2}$ (RD) and
$\bar{w}^{4}/u^{4}$ (MD) follow from the approximation that the
modes are deep inside the horizon: after emission, the metric
perturbation redshifts as $h^{2}\propto 1/a^{2}$, with
$a\propto\eta$ in RD and $a\propto\eta^{2}$ in MD. These expressions
apply when the emission and evaluation times are in the same era;
otherwise, e.g., for emission in RD and evaluation in MD, one
evaluates Eq.\,\eqref{eq:GW_Kernel} at $\min(\eta_{\rm eq},\,\eta_f)$
and redshifts the result to the evaluation time. To further simplify
the formula, we average the product of the Green's functions,
$\mathcal{G}(u,w')\,\mathcal{G}(u,w'')$, over one oscillation period
of the evaluation time $u$; in the UV regime $w',w''\gg 1$, this
average yields the factor
$\frac{1}{2}\cos(w_-)$.\footnote{Explicitly, deep inside the horizon,
$\mathcal{G}(u,w')\simeq \frac{w'}{u}\sin(u-w')$ in RD, so the
product of the two Green's functions contains
$\sin(u-w')\sin(u-w'')=\frac{1}{2}\left[\cos(w_-)
-\cos(2u-w'-w'')\right]$. The second term oscillates rapidly in $u$
and averages to zero over one period, leaving
$\mathcal{G}(u,w')\,\mathcal{G}(u,w'')\to
\frac{w'w''}{2u^{2}}\cos(w_-)$. The factor
$w'w''\simeq\bar{w}^{2}$ is absorbed into the transfer function in
\cref{eq:GW_Kernel}; the MD case follows analogously with
$\mathcal{G}(u,w')\simeq \frac{w'^{2}}{u^{2}}\sin(u-w')$.} The
kernel function encoding the two-point statistics of the source is
then
\begin{equation}
  K(\bar{w})
  \;=\;\frac{1}{2}
  \int_{-\infty}^{+\infty}\!dw_-\;\cos(w_-)\;
  \sum_{\lambda}\varepsilon^{ij}_{\lambda}\,
  \varepsilon^{\,*mn}_{\lambda}\;
  T_{ij,mn}\!\left(k,
    \frac{2\bar{w}+ w_-}{2k},
    \frac{2\bar{w}-w_-}{2k}\right),
  \label{eq:Decoherence_Kernel}
\end{equation}
We stress that the support of the integral \cref{eq:Decoherence_Kernel} is within the range of one correlation time of global string networks $\sim \xi\eta$, which results in the effective integration range of $|w_-|\lesssim k\xi\eta$. For the time scale longer than the correlation time, the UETC should decay and will not contribute significantly to the integral. For gauge string networks, the data-driven formalism needs to be modified to account for the contribution from long-lived gauge string loops, which we defer for a future study. 

\paragraph{Simulation input: the instantaneous GW emission spectrum.}
The kernel $K(u)$ has a direct physical meaning: it is the
instantaneous GW emission power spectrum of the network, which is the
quantity measured in lattice simulations. To make this connection,
consider the growth of the metric perturbation as the source evolves
continuously. Since the UETC has support only over time differences
shorter than the correlation time of the network for long strings,
or the loop lifetime for loops, the newly emitted contribution
correlates only with the emission in the recent past, and the $w_-$
integral in \cref{eq:Decoherence_Kernel} saturates over the full
support of the UETC. Both time scales are shorter than a Hubble
time, so the redshift over the correlated emission is negligible;
the subsequent propagation to a later evaluation time is restored by
the transfer function or simply by redshifting. The differential
two-point function of the metric perturbation then reduces to
\begin{equation}
    \frac{\partial }{\partial u}\sum_{\lambda}|h_{\lambda}(k, \eta)|^{2} \simeq \left(\frac{16\pi G}{k^{2}}\right)^{\!2}
  \,K(u)\,.
\end{equation}

The instantaneous GW radiation power spectrum, defined as the GW energy density emitted per unit time per logarithmic wavenumber interval, is therefore directly connected to the kernel function, 
\begin{equation}
\label{eq:Gamma_to_K}
  \frac{\partial\Gamma_{\rm GW}}{\partial \ln p}  =\frac{\partial^2\rho_{\rm GW}}{\partial t \partial\ln p}\simeq\frac{k^6}{64\pi^3 G a^3}\frac{\partial}{\partial u}\sum_\lambda|h_\lambda(k,\eta)|^2=\frac{4Gk^2}{\pi a^3}K(u),
\end{equation}
where $p=k/a$ is the physical wavenumber, and we used
$\partial\rho_{\rm GW}/\partial\ln p
=\frac{k^{5}}{64\pi^{3}Ga^{2}}\sum_\lambda|h_\lambda|^{2}$ together
with $\partial_t=(k/a)\,\partial_u$. In the second equality, the
sub-leading Hubble expansion term is neglected for UV modes
$k/a\gg H$, i.e., $k\eta\gg1$.

The simulation delivers a single quantity, the instantaneous GW emission spectrum $\partial\Gamma_{\rm GW}/\partial\ln p$. Following \cite{Gorghetto:2021fsn}, we factorize it into a dimensionless normalized shape $F_g(u)$, averaged within one Hubble time, and the total GW radiation power $\Gamma_{\rm GW}(t)$,
\begin{equation}
    \Gamma_{\rm GW}(t)=\int dp\frac{\partial\Gamma_{\rm GW}}{\partial p}, ~~~~\frac{\partial\Gamma_{\rm GW}}{\partial p}\equiv \frac{\Gamma_{\rm GW}(t)}{(a\eta)^{-1}}F_g(u),
    \label{eq:diff_Gamma}
\end{equation}
where the factorization implies $F_g(u)$ captures the momentum dependence only, and $(a\eta)^{-1}$ evaluated at $t$ is a normalization factor to convert $dp$ to $du$ such that $\int du F_g(u)=1$. The procedure to measure $F_g(u)$ is to perform lattice simulations of a complex scalar field $\phi$, evolve the metric perturbation $h_{ij}$ sourced by the transverse-traceless part of the stress-energy tensor $T^{\rm TT}_{ij}$, and extract the instantaneous GW emission spectrum from the time derivative of the $h_{ij}$ power spectrum within one Hubble time. In this work, we take $F_g(u)$ from Ref.~\cite{Gorghetto:2021fsn}, measured at $\mathcal{L}_f=5.8$ over $u\in[2.3,\,2000]$, and extrapolate it to higher modes as $F_g\propto u^{-2}$, characteristic of a kink-kink-collision-dominated spectrum. The amplitude of $\Gamma_{\rm GW}(t)$ can be estimated analytically, while its precise value is set by the form factor $\tilde{r}$ measured in simulations. Relating the GW power to the total radiated power through the branching ratio,
\begin{equation}
    \Gamma_{\rm GW}(t)= \tilde{r} \frac{G\mu_{\rm eff}^2}{f_a^2}\Gamma_a\simeq \tilde{r} \frac{G\mu_{\rm eff}^2}{f_a^2}\frac{\rho_{\rm str}}{t},
    \label{eq:Gamma_gw_rd}
\end{equation}
where $\tilde{r}\simeq0.26$~\cite{Gorghetto:2021fsn} characterizes the GW emission efficiency relative to axions. The second equality uses $\Gamma_a\simeq\Gamma_{\rm tot}=\rho_{\rm str}/t$, set by the scaling energy balance since global string loops decay predominantly into axions, where $\rho_{\rm str}=\mu_{\rm eff}/L_c^2$ is the long-string energy density and $L_c=a\xi\eta$ is the physical correlation length. The overall normalization is therefore fixed by the total energy processed by the scaling network, while the spectral shape is set by $F_g$, dominated by loops and small-scale structure.

\paragraph{Data-driven UV spectrum.}
After obtaining the two quantities, $F_g(u)$ and $\Gamma_{\rm GW}(t)$, we use \cref{eq:Gamma_to_K,eq:diff_Gamma} to do the mapping between the instantaneous radiation power spectrum and the kernel function,
\begin{equation}
    K(u)=\frac{\pi a^3}{4Gk^2} \frac{\partial\Gamma_{\rm GW}}{\partial \ln p}=\frac{\pi a^3\Gamma_{\rm GW}(t)}{4Gk^2}\, uF_g(u)\,.
    \label{eq:K_from_Fg}
\end{equation}
The present-day GW spectrum follows from \cref{eq:OmGW_def},
\begin{equation}
   \Omega_{\rm GW,0}(k)\;=\;\frac{1}{12}\left(\frac{k}{a_0 H_0}\right)^{\!2}
  \frac{k^{3}}{2\pi^{2}}\left(\frac{a_f}{a_0}\right)^2\sum_{\lambda}|h_{\lambda}(k, \eta_f)|^{2},
  \label{eq:Omega_from_h2}
\end{equation}
where the term $(a_f/a_0)^2$ redshifts the two-point function from the decay time to today, valid for subhorizon modes since they have already re-entered the horizon and behave as radiation. Substituting the kernel~\cref{eq:K_from_Fg} and the total power~\cref{eq:Gamma_gw_rd} into \cref{eq:GW_Kernel,eq:GW_Kernel_MD}, and then into \cref{eq:Omega_from_h2}, we obtain the final results of the UV spectrum:
\begin{equation}
    \Omega_{\rm GW,0}(k)=\frac{16\pi^2(G\pi f_a^2)^2}{3}\tilde{r}\int_{u_i}^{u_f} dw \ln^3\left(\frac{a \xi w}{k\delta}\right) F_g(w)\times \begin{dcases}
   \frac{\Omega_r}{\xi^{2}}
    \left(\dfrac{g_{*s}(T_f)}{g_{*s,0}}\right)^{\!-1/3},
    & \text{RD}\,\\[6pt]
    \frac{3}{32\,\xi^{2}}\,w^2\Omega_m^{2}(a_0H_0 k^{-1})^{2}\,,
    & \text{MD}\,
  \end{dcases},
\label{eq:Omega_uv_dd}
\end{equation}
where $\xi$ is the scaling parameter of the network and the string core size is $\delta = \sqrt{2/\lambda}\,f_a^{-1}$ (with the self-coupling set to $\lambda = 1$).\footnote{Adopting a cusp-dominated tail $F_g\propto w^{-4/3}$ instead of the kink-dominated $w^{-2}$ allows a comparison with the harmonic-method loop spectrum of \cite{Chang:2021afa}. We recover the same $k^{-1/3}$ low-frequency slope but a smaller amplitude. After the investigation, we found that the difference may arise from the transfer function; using the RD function \cref{eq:GW_Kernel} in place of the MD one \cref{eq:GW_Kernel_MD} reproduces their result.}

\section{Gravitational Wave Spectra}
\label{sec:GWSpectra}
In this section, we present the GW spectrum produced by a scaling
global string network during the radiation-dominated and matter-dominated era, combining
the analytical IR results from the USM (\cref{sec:gwspectraldensity})
with the data-driven UV method (\cref{sec:uvregime}). We first
summarize the observational landscape, then present the full spectrum
for benchmark parameters and discuss its dependence on the axion
mass~$m_a$ and the symmetry-breaking scale~$f_a$.

\subsection{Observational Landscape}
\label{sec:observations}
The stochastic GW background from cosmic strings spans a broad frequency range, from nanohertz to kilohertz. We summarize the relevant observational constraints and projected sensitivities in Table~\ref{tab:GW_missions}.
\begin{table}[htbp]
\centering
\renewcommand{\arraystretch}{1.2}
\begin{tabular}{llll}
\toprule
\textbf{Frequency band} & \textbf{Method/Experiment}
  & \textbf{Status} & \textbf{Refs.} \\
\midrule
\multicolumn{4}{l}{\textit{Integrated constraints}} \\
\quad $\Delta N_{\rm eff}$ (BBN)
  & $\int\Omega_{\rm GW}\,d\ln f$ & Current
  & \cite{Cyburt:2015mya} \\
\quad $\Delta N_{\rm eff}$ (Planck+BAO)
  & $\int\Omega_{\rm GW}\,d\ln f$ & Current
  & \cite{Planck:2018vyg} \\
\midrule
\multicolumn{4}{l}{\textit{Indirect probes}} \\
\quad $\sim\!10^{-18}$--$10^{-16}$\,Hz
  & CMB $B$-modes (Planck, BICEP/Keck) & Current
  & \cite{Planck:2018vyg,BICEP:2021xfz} \\
\quad $\sim\!10^{-18}$--$10^{-16}$\,Hz
  & CMB $B$-modes (LiteBIRD) & Projected
  & \cite{Hazumi:2019lys,LiteBIRD:2022cnt} \\
\quad $\sim\!10^{-15}$--$10^{-9}$\,Hz
  & Spectral distortions (PIXIE, Voyage 2050) & Projected
  & \cite{Kogut:2019vqh,Chluba:2019nxa} \\
\midrule
\multicolumn{4}{l}{\textit{Direct and astrometric detection}} \\
\quad $10^{-9}$--$10^{-7}$\,Hz
  & PTAs (NANOGrav, EPTA, PPTA, CPTA) & Current
  & \cite{NANOGrav:2023hvm,EPTA:2023sfo,Reardon:2023gzh,Xu:2023wog} \\
\quad $10^{-9}$--$10^{-7}$\,Hz
  & PTAs (SKA) & Projected
  & \cite{Weltman:2018zrl} \\
\quad $10^{-8}$--$10^{-6}$\,Hz
  & Astrometry (Gaia, THEIA) & Projected
  & \cite{Garcia-Bellido:2021zgu} \\
\quad $10^{-6}$--$10^{-3}$\,Hz
  & Space interf.\ ($\mu$Ares) & Projected
  & \cite{Sesana:2019vho} \\
\quad $10^{-4}$--$10^{-1}$\,Hz
  & Space interf.\ (LISA) & Projected
  & \cite{LISA:2017pwj} \\
\quad $10^{-2}$--$10^{1}$\,Hz
  & Atom interf.\ (AION, AEDGE) & Projected
  & \cite{Badurina:2021rgt,AEDGE:2019nxb} \\
\quad $10^{-2}$--$10^{2}$\,Hz
  & Space interf.\ (DECIGO, BBO) & Projected
  & \cite{Yagi:2011wg,Crowder:2005nr} \\
\quad $10^{0}$--$10^{2}$\,Hz
  & Ground interf.\ (aLIGO/Virgo/KAGRA) & Current
  & \cite{KAGRA:2021kbb} \\
\quad $10^{0}$--$10^{4}$\,Hz
  & Ground interf.\ (ET, CE) & Projected
  & \cite{Punturo:2010zz,Reitze:2019iox} \\
\bottomrule
\end{tabular}
\caption{Summary of observational constraints and projected
  sensitivities relevant to the stochastic GW background from cosmic
  strings. Frequency entries for indirect probes (CMB $B$-modes,
  spectral distortions) denote the approximate cosmological scales
  probed rather than instrumental frequency bands. ``Current'' denotes
  operating experiments or recently published results; ``Projected''
  denotes future sensitivity.}
\label{tab:GW_missions}
\end{table}

The integrated GW energy density is bounded by constraints on additional relativistic degrees of freedom~\cite{Maggiore:1999vm}:
\begin{equation}
\int_{f_{\rm min}}^{\infty} \frac{df}{f}\, \Omega_{\rm GW}(f)\, h^2 
\;\leq\; 5.6 \times 10^{-6}\, \Delta N_{\rm eff}\,.
\label{eq:Neff_bound}
\end{equation}
For Planck 2018 + BAO, one finds $\Delta N_{\rm eff} \lesssim 0.3$ at 95\% confidence level (C.L.).

\subsection{GW Spectrum of Long Gauge Strings}
\label{app:gaugestrings}
Before presenting the global-string spectrum, we validate the USM IR
formalism on long gauge strings, the constant-tension limit of our
calculation, for which independent Nambu-Goto simulations are
available. The gauge-string spectrum follows from the global-string
results of \cref{subsec:longstrings} by replacing the running
effective tension with a constant, $\mu_{\rm eff}(\eta)\to\mu=\pi
f_a^2$: the energy is localized within the string core, so there is
no long-range field and hence no infrared logarithm, in contrast to
global strings. We then compare
the analytical long-string IR spectra against the long-string spectra
measured in Nambu-Goto simulations~\cite{CamargoNevesdaCunha:2022mvg}.
As throughout, we set the wiggliness parameter to $\alpha=1$,
corresponding to featureless segments.

We separate the contributions sourced during the radiation- and
matter-dominated eras, corresponding to $\eta_f=\eta_{\rm eq}$ and
$\eta_f=\eta_0$, respectively, and directly present the analytical
results here:
\begin{equation}
  \Omega_{\rm GW,0}^{\rm(I)}(k)
  \;=\;\frac{128\,(G\mu)^{2}}{9 \,{a_0^{2}H_0^{2} \eta_0^{2}}}\;\xi^{-1}\;
  F_{12,12}(v)\;
  \;\times \begin{cases} \frac{1}{3}
  (k\eta_f)^3\,,
  \qquad 1 < k\eta_{\rm eq}\,,\ \eta_f < \eta_{\rm eq} \,,\\
  \\
  \underline{\frac{3}{2} \left(\frac{\eta_f}{\eta_0}\right)^2 \cdot k\eta_f} \,, 
  \qquad k\eta_{\rm eq} < 1 < k\eta_0\,,\ \eta_f < \eta_{\rm eq}\,,\\
  \\
  \underline{\frac{1}{6} \left(\frac{\eta_0}{\eta_f}\right)^2 \cdot (k\eta_f)^5} \,,
  \qquad k\eta_{0} < 1\,,\ \eta_{\rm eq}<\eta_f  \,,\\
  \\
  \frac{1}{12} \left(\frac{\eta_f}{\eta_0}\right)^2 \cdot k\eta_f \,,
  \qquad k\eta_f< 1<k\eta_{0} \,,\ \eta_{\rm eq}<\eta_f \,, \\
  \end{cases}
  \label{eq:OmGW_I_gauge}
\end{equation}

\begin{equation}
  \Omega_{\rm GW,0}^{\rm(II)}(k)
  \;=\;\frac{128\,(G\mu)^{2}}{3}\;\xi^{-1}\;
  F_{12,12}(v)\;(k\eta_f)^{3}
  \;\times\;
  \begin{dcases}
    \dfrac{\Omega_r}{9}
    \left(\dfrac{g_{*s}(T_f)}{g_{*s,0}}\right)^{\!-1/3},
    & 1 < k\eta_f\,,\ \eta_f<\eta_{\rm eq}\,,\\[6pt]
    \dfrac{\Omega_m^{2}\,a_0^{2}H_0^{2}\,\eta_f^{2}}{32\,}\,,
    & 1 < k\eta_f\,,\ \eta_f>\eta_{\rm eq}\,,
  \end{dcases}
  \label{eq:OmGW_II_gauge}
\end{equation}

\begin{equation}
  \Omega_{\rm GW,0}^{\rm(III),\rm IR}(k)
  \;=\;\frac{128\,(G\mu)^{2}}{3}\;\xi^{-4}\;
  F_{12,12}(v)
  \;\times\;
  \begin{dcases}
    \dfrac{\Omega_r}{9}
    \left(\dfrac{g_{*s}(T_f)}{g_{*s,0}}\right)^{\!-1/3},
    & \xi^{-1} < k\eta_f\,,\ \eta_f<\eta_{\rm eq}\,,\\[6pt]
    \dfrac{\Omega_m^{2}\,a_0^{2}H_0^{2}\,\eta_f^{2}}{32} \frac{\xi^{-2}}{k^2\eta_f^2}\,,
    & k\eta_{\rm eq} < \xi^{-1} < k\eta_f\,,
  \end{dcases} \, . \,
  \label{eq:OmGW_III_gauge}
\end{equation}

\begin{figure}[htbp]
\centering
\includegraphics[width=0.75\linewidth]{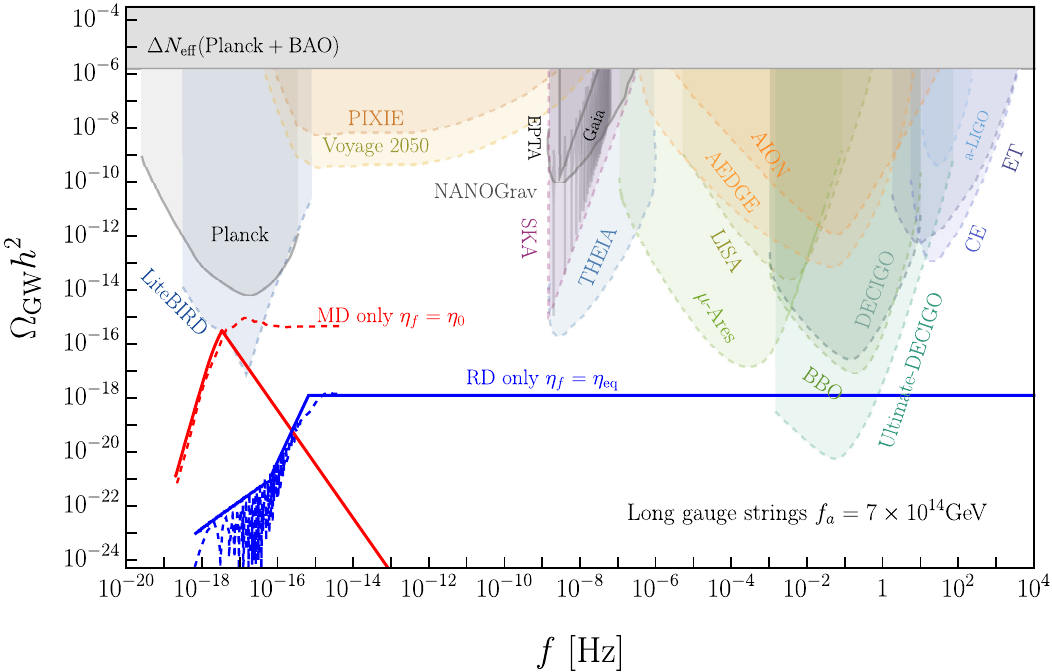}
\caption{Long-string only GW spectrum from a scaling Nambu-Goto string network with $f_a = 7 \times 10^{14}\,\mathrm{GeV}$. \textbf{Solid lines:} UETC calculation with IR contribution only (cutoff at $k\xi\eta = 1$). 
\textbf{Dashed lines:} Simulation results from Ref.~\cite{CamargoNevesdaCunha:2022mvg}. Blue curves show the RD-era contribution and red curves show the MD-era contribution. Shaded regions indicate sensitivity curves for current and future GW observatories. The gray band shows the Planck+BAO $\Delta N_{\rm eff}$ bound.
}
\label{fig:gauge_spectrum}
\end{figure}

Fig.~\ref{fig:gauge_spectrum} shows the GW spectrum from a scaling Nambu-Goto string network (long-string contributions only) with $f_a = 7 \times 10^{14}\,\mathrm{GeV}$, chosen so that $G\mu \simeq 10^{-8}$ matches the Nambu-Goto simulation of Ref.~\cite{CamargoNevesdaCunha:2022mvg}. The RD (blue) and MD (red) contributions are shown separately, computed with the USM/UETC formalism; the solid curves include only the IR contribution, with the UETC cutoff at $k\xi\eta = 1$ (i.e., wavelengths larger than the correlation length).

We stress that the blue and red curves are obtained by numerically
evaluating the GW master formula~\eqref{eq:OmGW_master} with the USM
source UETC, without invoking any piecewise approximation. Their
spectral shapes can nonetheless be understood analytically from the
regime formulas
\cref{eq:OmGW_I_gauge,eq:OmGW_II_gauge,eq:OmGW_III_gauge}, evaluated
at $\eta_f=\eta_{\rm eq}$ for the RD contribution and $\eta_f=\eta_0$
for the MD contribution, as we now describe. 

For the RD contribution (blue solid), the analytic formulas are
evaluated at $\eta_f=\eta_{\rm eq}$. Strictly, the piecewise
conditions in these equations are written for $\eta_f<\eta_{\rm eq}$,
whereas the RD-sourced spectrum corresponds to the boundary value
$\eta_f=\eta_{\rm eq}$; the formulas nonetheless provide a good
approximation there. The branch that applies is fixed by comparing
$k\eta_{\rm eq}$ to the two characteristic scales $1$ and $\xi^{-1}$,
giving three spectral shapes from low to high frequency:
\begin{itemize}
\item $k\eta_{\rm eq}<1$ (regime~I): superhorizon modes at
  $\eta_{\rm eq}$ that re-enter during MD, from the underlined branch
  of \cref{eq:OmGW_I_gauge}, giving $\Omega_{\rm GW,0}\propto k$;
\item $1<k\eta_{\rm eq}<\xi^{-1}$ (regime~II): each correlation volume
  acts as an uncorrelated source, from the RD branch of
  \cref{eq:OmGW_II_gauge}, giving the white-noise
  $\Omega_{\rm GW,0}\propto k^{3}$;
\item $k\eta_{\rm eq}>\xi^{-1}$ (regime~III): the IR contribution with
  the UETC truncated at $u_{f,*}=\xi^{-1}$, from the RD branch of
  \cref{eq:OmGW_III_gauge}, giving a plateau
  $\Omega_{\rm GW,0}\propto k^{0}$, as for long global strings in
  \cref{subsec:longstrings}.
\end{itemize}

For the MD contribution (red solid), the analytic formulas are
evaluated at $\eta_f=\eta_0$, selecting the branches with
$\eta_f>\eta_{\rm eq}$. The branch that applies is fixed by comparing
$k\eta_0$ to the two characteristic scales $1$ and $\xi^{-1}$, giving
three spectral shapes from low to high frequency:
\begin{itemize}
\item $k\eta_0<1$ (regime~I): superhorizon modes that remain frozen
  at the present time, from the underlined branch of
  \cref{eq:OmGW_I_gauge}, giving $\Omega_{\rm GW,0}\propto k^{5}$;
\item $1<k\eta_0<\xi^{-1}$ (regime~II): each correlation volume acts
  as an uncorrelated source, from the MD branch of
  \cref{eq:OmGW_II_gauge}, giving the white-noise
  $\Omega_{\rm GW,0}\propto k^{3}$;
\item $k\eta_0>\xi^{-1}$ (regime~III): the IR contribution with the
  UETC truncated at $u_{f,*}=\xi^{-1}$, from the MD branch of
  \cref{eq:OmGW_III_gauge}, giving
  $\Omega_{\rm GW,0}\propto k^{-2}$.
\end{itemize}

For comparison, the dashed lines show the results of Nambu-Goto
string simulations~\cite{CamargoNevesdaCunha:2022mvg}. Our
semi-analytic UETC calculation reproduces the simulated spectra
across both the RD and MD contributions, validating the USM as a
reliable description of the long-string IR spectrum. This benchmark
justifies applying the same formalism to global strings.

In \cref{fig:gauge_spectrum}, the long-string UV spectra from Nambu-Goto simulations at both the RD and MD are also shown, which corresponds to the regime III frequency range $k\gtrsim \xi^{-1}/\eta_{\rm eq}$\,(RD) and $k\gtrsim \xi^{-1}/\eta_{0}$\,(MD). The discrepancy between the semi-analytic UETC result (red solid line) and the simulation result (red dashed line) in the UV spectrum for MD is because 1. the UETC result is IR-contribution only, while the simulation result contains both IR and UV contributions; 2. the UV-contribution dominates over the IR-contribution.

\begin{figure}[t!]
\centering
\includegraphics[width=0.75\linewidth]{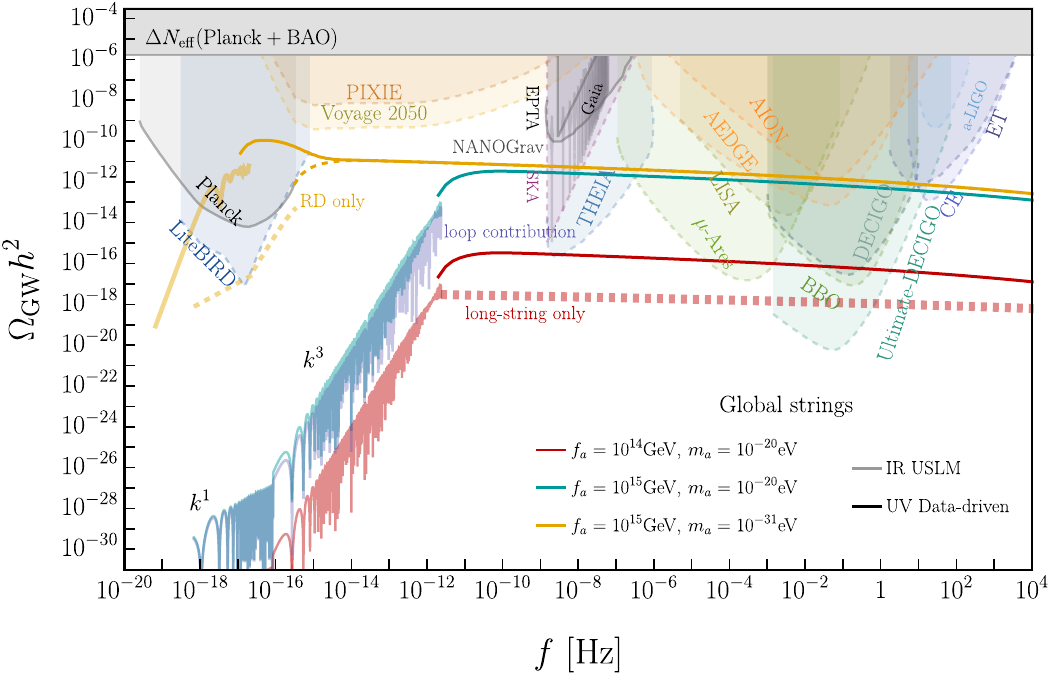}
\caption{GW spectrum from a scaling global string network with
  $f_a = 10^{14},\,10^{15}\,$GeV.
  \textbf{Thin opaque lines:} UV part of the spectrum from the data-driven approach. \textbf{Thick transparent lines:} IR part from the USM. Different colors correspond to different values of~$m_a$ (see legend), which set the network decay time via $H(\eta_f) = m_a$. 
  For $m_a=10^{-31}\,$eV case, the corresponding RD-only contribution is shown in dashed line, while for $m_a=10^{-20}\,$eV case, the corresponding long-string only contribution is shown in dashed line.
  Shaded regions show the projected sensitivities of current and future GW observatories
  (Table~\ref{tab:GW_missions}), and the gray band denotes the Planck+BAO $\Delta N_{\rm eff}$ bound~\eqref{eq:Neff_bound}.}
\label{fig:global_spectrum}
\end{figure}

\subsection{GW Spectrum of Global String Networks}
\label{sec:global_spectrum}
We now present the main results of this work: the GW spectrum from a
scaling global string network, and its confrontation with current and
future GW experiments. The infrared part is obtained by numerically
evaluating the GW master formula with the UETC of the unconnected
segment and loop model, which the analytical results of
\cref{sec:gwstrings} reproduce in each regime, and the ultraviolet part
from the data-driven method of \cref{sec:uvregime}. We focus on axion
models in which the Peccei-Quinn symmetry is broken after inflation.

The results are summarized in \cref{fig:global_spectrum}, where the two
parts are labeled in the legend: the IR spectrum is the rapidly
oscillating rising portion at low frequencies, and the UV spectrum the
smooth plateau at higher frequencies. We show three benchmarks:
$(f_a,\,m_a)=(10^{14}\,{\rm GeV},\,10^{-20}\,{\rm eV})$ and
$(10^{15}\,{\rm GeV},\,10^{-20}\,{\rm eV})$, for which the network decays
deep in the RD era, and
$(10^{15}\,{\rm GeV},\,10^{-31}\,{\rm eV})$, for which it decays during
the MD era. The first two differ only in $f_a$ and thus illustrate how
the amplitude scales with the symmetry-breaking scale at fixed spectral
shape.

Several general features of the spectrum are apparent. First, the
spectrum rises in the IR and flattens into a plateau in the UV. The
smooth transition between the IR and UV
spectra demonstrates that the USLM captures the correct long-wavelength
physics and provides a reliable analytical method for deriving the IR
spectrum. The visible mismatch at the matching point is about a factor
of a few, which can be attributed to the theoretical and simulation
uncertainties of the string network.

Second, the network terminates when the axion mass becomes comparable
to the Hubble rate, $m_a\simeq H$, at which point the domain-wall
dynamics dominates. This determines the final time of the network
$\eta_f$, and hence the frequency at which the rising IR spectrum turns
over into the UV plateau, $f_{*}\simeq 2\times 2\pi/(\xi\eta_f)$.
A lighter axion corresponds to a later decay and thus a lower turnover
frequency, shifting the spectrum to the left in
\cref{fig:global_spectrum}.

Third, compared with the constant-tension case of gauge strings, the
amplitude is enhanced by powers of the logarithm $\mathcal{L}_f$,
reaching $\mathcal{L}_f^{3}$ for the UV plateau.
Fourth, the UV plateau is not exactly flat. The frequency dependence
can be seen from~\cref{eq:Omega_uv_dd}, which gives
$\Omega_{\rm GW}\propto\ln^{3}(A/k)$ with $A$ a constant, a mild
logarithmic tilt that breaks exact scale invariance.

We now present more features based on the three benchmarks shown in
\cref{fig:global_spectrum}. We begin with
$(f_a,\,m_a)=(10^{14}\,{\rm GeV},\,10^{-20}\,{\rm eV})$. The
network terminates when $m_a\simeq H$, which for this axion mass occurs
deep in the RD era, at a temperature $T_f\simeq 6\,$keV. The IR part of
the spectrum exhibits a clear $k^{3}$ power law, reflecting the
white-noise nature of the string source, and includes both the
long-string and loop contributions computed within the USLM. The thick dashed
line shows the long-string contribution alone in the
sub-correlation regime~(III), which is sourced at earlier times
$\eta<\eta_f$. For $f\gtrsim10^{-10}\,$Hz it lies one to two orders of
magnitude below the UV spectrum from loops obtained with the
data-driven method.

The second benchmark,
$(f_a,\,m_a)=(10^{15}\,{\rm GeV},\,10^{-20}\,{\rm eV})$, differs
from the first only in the symmetry-breaking scale. Since the amplitude
scales as $f_a^{4}$, the whole spectrum is rescaled upward by about four
orders of magnitude, while its shape and turnover frequency remain
unchanged. For this benchmark we also show the loop contribution to the
IR spectrum separately, as the transparent line, and it is close to the
long-string one, confirming that the two contributions are comparable.
The solid oscillating curve is their sum. At the lowest frequencies, the
IR modes re-enter the horizon after $\eta_{\rm eq}$, and the spectrum
follows the regime~(I) scaling $\Omega_{\rm GW}\propto k$, as indicated
in the figure. For this value of $f_a$, the UV plateau spans the
frequency bands of SKA, LISA, DECIGO, BBO, ET, and CE, and its amplitude
$\Omega_{\rm GW}h^{2}\sim 10^{-12}-10^{-11}$ passes through the
NANOGrav sensitivity band, placing it within reach of most planned
detectors.

The third benchmark,
$(f_a,\,m_a)=(10^{15}\,{\rm GeV},\,10^{-31}\,{\rm eV})$, has a much
smaller axion mass, so the network survives into the MD era, terminating
at $z_f\simeq24$, corresponding to $T_f\simeq6\,$meV, and the
turnover between the IR and UV spectra moves to a much lower frequency,
$f_{*}\simeq10^{-19}\,$Hz. For this benchmark we also show the RD-only
contribution as the dashed line, obtained by integrating the source only
up to $\eta_{\rm eq}$. Above $f\simeq10^{-15}\,$Hz the two curves merge,
showing that the high-frequency part of the spectrum is dominated by the
modes sourced at earlier times. The IR part of the RD-only contribution
displays the $k^{3}$ and $k$ power laws discussed above. For the full
contribution, the network decays at $\eta_f>\eta_{\rm eq}$, and
\cref{eq:Omega_gw_I,eq:Omega_gw_II} predict three IR slopes from low to
high frequency: $\Omega_{\rm GW}\propto k^{5}$ for modes that are still
superhorizon today, $k\eta_0<1$, then $\Omega_{\rm GW}\propto k$ in the
window $k\eta_f<1<k\eta_0$, and finally the white-noise
$\Omega_{\rm GW}\propto k^{3}$ for $1<k\eta_f<\xi^{-1}$. Since
$\eta_f\simeq0.2\,\eta_0$ for this benchmark, the latter two windows
span less than a decade in frequency each, so they are not resolved in
\cref{fig:global_spectrum}.

Finally, we derive the resulting constraint on the axion parameter
space. For ultralight axion masses,
$2.25\times10^{-28}\,\text{eV}\lesssim m_a\lesssim 10^{-23}\,\text{eV}$, the IR
spectrum enters the CMB $B$-mode window
($f\sim 10^{-18}$-$10^{-15}\,$Hz), where a stochastic GW background
sources tensor perturbations at the last-scattering surface and
contributes to the $B$-mode power spectrum on large angular
scales~\cite{Seljak:1996gy,Kamionkowski:1996zd}. This provides the
tightest constraint on the model. The current Planck $B$-mode bound,
corresponding to $\Omega_{\rm GW}h^{2}\sim 10^{-13}$ at these
frequencies, yields
\begin{equation}
  f_a
  \;\lesssim\;2.75\times10^{15}\,\text{GeV}
  \left(\frac{m_a}{10^{-27}\,\text{eV}}\right)^{\!0.38},
  \qquad
  2.25\times10^{-28}\,\text{eV}\lesssim m_a
  \lesssim 10^{-23}\,\text{eV}\,,
  \label{eq:fa_Bmode}
\end{equation}
where the power-law scaling reflects the $k^{3}$ behavior of the
regime~(II) spectrum in the Planck-sensitive frequency range.
LiteBIRD~\cite{Hazumi:2019lys,LiteBIRD:2022cnt} will improve this bound
by roughly an order of magnitude in $\Omega_{\rm GW}h^{2}$,
corresponding to a factor of $\sim 1.8$ in~$f_a$. We emphasize that
this constraint follows from the IR spectrum sourced during RD, which is
accessible only once the IR regime is included in the calculation.
For heavier axions
the turnover moves to higher frequencies and the $B$-mode window probes
only the slowly growing regime~(I) tail, so that $B$-mode searches lose
sensitivity. In that case the relevant constraints come instead from
direct GW observations, since the UV plateau enters the PTA band and
the sensitivity ranges of planned interferometers, as discussed above.
Requiring the plateau not to exceed the stochastic signal observed by
PTAs (the EPTA constraint is chosen for deriving the bound), $\Omega_{\rm GW}h^{2}\lesssim10^{-9}$ in the nanohertz band,
gives
\begin{equation}
  f_a\;\lesssim\;3.2\times10^{15}\,\text{GeV}\,,
  \qquad
  m_a\;\lesssim\;10^{-15}\,\text{eV}\,,
  \label{eq:fa_PTA}
\end{equation}
where the mass range is set by requiring the turnover to lie below the
PTA band. SKA will improve this to
$f_a\lesssim 1.9\times10^{14}\,$GeV. 

For comparison, we summarize the existing constraint on global string tension and axion mass by classifying them into model-independent ones (via only gravitational effect) and model-dependent ones (relying on couplings between the scalar field that forms the string and the Standard Model).

\paragraph{Model-independent constraints:} The reheating temperature $T_{\rm RH}\gtrsim  f_a$ for a post-inflationary second-order phase transition to happen. The latest B-mode measurement by BICEP/Keck telescopes (BK18) combined with publicly available WMAP and Planck data \cite{BICEPKeck:2022mhb}, place constraint on the tensor-to-scalar ratio $r<0.036$ (95\% C.L.) and further infer that the reheating temperature $T_{\rm RH}\lesssim 1.4\times 10^{16}\,$GeV. In order for a successful string formation in the post-inflationary scenario, $f_a\lesssim 1.4 \times 10^{16}\,$GeV.

The CMB angular power spectrum $C_\ell$ can also place constraints on the existence of a cosmic string network through its density inhomogeneity on top of the cosmic background, $G\mu\lesssim1.1\times 10^{-7}$
\cite{Charnock:2016nzm}, which gives $f_a\lesssim2.1\times 10^{14}\,$GeV if the string network can survive until around or after recombination. 

The large scale structure observation can also place constraints on axion string networks because the string network induces isocurvature perturbation, even though sub-fractional \cite{Gorghetto:2025uls}. The best existing constraint is $f_a\lesssim 3\times 10^{14}\,$GeV for $m_a\sim 10^{-24}-10^{-23}\,$eV. 

By considering only the contribution from relativistic axions to the $\Delta N_{\rm eff}$ at BBN, Ref.\,\cite{Gorghetto:2021fsn} constrains $f_a\lesssim 9\times10^{14}\,$GeV (assuming scaling violation). If the string network persists after recombination, the $\Delta N_{\rm eff}<0.34$ constraint at CMB \cite{Planck:2018vyg} is even tighter, $f_a\lesssim 3\times 10^{14}\,$GeV.  

\paragraph{Model-dependent constraints:} Ref.\,\cite{Benabou:2023ghl} further consider the energy injection to Standard Model thermal bath by radial mode emission from string decay. They obtain a stronger constraint on $f_a\lesssim 1\times 10^{14}\,$GeV$\sqrt{\frac{0.33}{c}\frac{1}{\mathcal{B}}\frac{25}{N_s(z_*)}}$ (BBN) for $m_a\lesssim 2\times 10^{-23}\,$eV, and $f_a\lesssim 1.06\times 10^{12}\,$GeV$\sqrt{\frac{0.33}{c}\frac{1}{\mathcal{B}}\frac{30}{N_s(z_{\rm inj})}}$ (CMB), where $c$ is radial mode emission rate coefficient, $\mathcal{B}$ is the branching ratio for the radial mode decay to Higgs boson, $N_s$ is the average long string number per Hubble patch, $z_*$ is some redshift when the bound is evaluated, $z_{\rm inj}=600$.

\section{Conclusions}
\label{sec:Conclusions}
In this work, we computed the gravitational wave spectrum from scaling global string networks across infrared ($k\lesssim\ell_c^{-1}$) and ultraviolet ($k\gtrsim\ell_c^{-1}$) scales, providing for the first time the complete spectrum. In the IR, we derived the GW spectrum analytically using the unequal-time correlator (UETC) of the string stress-energy tensor. A key theoretical development is the extension of the unconnected segment
model (USM) to incorporate the loop contribution. The network parameters, correlation length and RMS velocity, are determined by the velocity-dependent one-scale (VOS) model in the scaling
regime. In the UV, where sub-correlation-length structure dominates the source, we developed a data-driven method that uses the instantaneous radiation power spectrum measured in lattice simulations as input.

For the IR spectrum, we identified three spectral regimes, classified by the horizon and correlation scales at the network decay time: superhorizon (I), subhorizon super-correlation (II), and sub-correlation (III). In regime~I, the slope is set by the epoch of horizon re-entry, with $\Omega_{\rm GW}\propto k^{3}$, $k$, or $k^{5}$ for modes re-entering during RD, re-entering during MD, or remaining superhorizon today. Regime~II exhibits the white-noise spectrum $\Omega_{\rm GW}\propto k^{3}$. In regime~III, the IR contribution flattens to a plateau, up to a mild logarithmic tilt from the running effective tension, for GWs sourced during RD, and falls as $k^{-2}$ for those sourced during MD. For realistic parameters, the loop and long-string contributions to the IR spectrum are comparable.

The data-driven method connects the UETC formalism to the instantaneous radiation power spectrum measured in lattice simulations~\cite{Gorghetto:2021fsn}. The central object is a kernel function, derived from the UETC by averaging the Green's functions over an oscillation period, which encodes the two-point statistics of the source and is directly related to the instantaneous GW emission spectrum. Given this kernel, the UV GW spectrum follows by propagating the emitted waves to the present with the appropriate cosmological transfer function. We observed that the IR (semi-analytical) and UV (data-driven) spectra match smoothly, providing a non-trivial consistency check of the framework, with a residual offset of a factor of a few attributable to theoretical and simulation uncertainties of the string network.

Compared to gauge strings, the global string GW spectrum exhibits three key differences: (i)~the amplitude is enhanced by powers of the logarithm $\mathcal{L}_f$: the IR long-string contribution scales as $\mathcal{L}_f^{2}$ from the effective tension, and the UV plateau as $\mathcal{L}_f^{3}$, with one additional power from the total radiation power of the scaling network; (ii)~the turnover frequency between the IR and UV spectra is set by the axion mass via $f_{*}\simeq 4\pi(\xi\eta_f)^{-1}$ with $H(\eta_f)\simeq m_a$, making the spectrum a direct probe of the axion
parameter space; and (iii)~the UV spectral shape acquires logarithmic corrections from the $k$-dependent effective tension that break exact scale invariance.

We confronted the predicted spectra with current constraints and projected sensitivities across multiple frequency bands. For ultralight axion masses $2.25\times10^{-28}\lesssim m_a/\text{eV}\lesssim 10^{-23}$, the IR spectrum sourced during RD enters the CMB $B$-mode window, yielding the constraint $f_a\lesssim 2.75\times10^{15}(m_a/10^{-27}\,\text{eV})^{0.38}\,$GeV. This constraint is inaccessible to UV-only approaches. For heavier axions, the UV plateau enters the PTA band and the sensitivity ranges of planned interferometers such as LISA, DECIGO, BBO, ET, and CE, where the mild logarithmic tilt of the spectral shape becomes a distinctive signature of global strings.
 
Several directions remain for future work. A Bayesian analysis of the predicted spectral shape against the NANOGrav data, properly accounting for the supermassive black hole binary (SMBHB) foreground, would map out the excluded region in the $(m_a,\,f_a)$ plane. The USM framework can be extended to incorporate the loop contribution for gauge strings. The formalism also provides the foundation for computing the string contribution to CMB temperature and polarization anisotropies and to large-scale structure, and for studying the impact of deviations from exact scaling on the GW spectrum.

\section*{Acknowledgments}  The authors would like to thank Junwu Huang, Sergey Sibiryakov, Pierre Sikivie, and Tanmay Vachaspati for useful discussions. The work of SV is supported by the Kavli Institute for Cosmological Physics at the University of Chicago. CL and WX are supported in part by the U.S. Department of Energy under grant DE-SC0022148 at the University of Florida.
FY was supported by the NSF Grant Number PHY-2412701. 


\appendix

\newpage 
\section{Cosmic Strings and their Correlators}\label{app:UETC}

In this appendix we derive the UETC (and subsequent ETC) of the cosmic string tensors and use them to find the gravitational wave intensity of the corresponding networks. Since the core difference between global and gauge strings is their tension, we are going to perform their analysis in parallel, pointing out any differences in the process.

\subsection{String trajectory in USM model\label{app:string_trajectory}}
The Fourier transform of the stress-energy tensor is
\begin{equation}
\label{eq:Theta_FT}
  T_{\mu\nu}({\bf k},\eta) \; = \; \int d^3{\bf x} \, e^{-i{\bf k}\cdot{\bf x}} \, T_{\mu\nu}({\bf x},\eta) \, .
\end{equation}
To evaluate this transformation from $T_{\mu\nu}({\bf x},\eta)$, we must model the string trajectory. Following the unconnected segment model (USM) approach~\cite{Pen:1997ae,Pogosian:1999np,Avgoustidis:2012gb}, widely used in the literature, we approximate the string network as a collection of straight segments. This approximation is valid when the scale of interest $k^{-1}$ is much larger than the scale of small-scale structure (wiggles) on the strings.

For a single straight segment of comoving length $\ell = \xi\eta$, moving with constant velocity $v$ in the transverse direction, the trajectory is
\begin{equation}
    \label{eq:trajectory}
    {\bf x}_s(\sigma,\eta) \; = \; {\bf x}_0+v\eta \hat{\bf v}+\sigma \hat{\bf x} \,,
\end{equation}
where $\sigma \in [-\ell/2, \ell/2]$ parametrizes the position along the segment, $\mathbf{x}_0$ is the initial position of the segment center, $\hat{\mathbf{x}}=(\sin\theta\cos\varphi,\sin\theta\sin\varphi,\cos\theta)$ is the unit vector along the string direction, and $\hat{\mathbf{v}}$ is the velocity unit vector satisfying $\hat{\mathbf{v}}\perp\hat{\mathbf{x}}$. Without loss of generality, we align the wavevector along the $z$-axis, $\mathbf{k}=k\hat{\mathbf{z}}$. Performing the Fourier transform \eqref{eq:Theta_FT} with the trajectory \eqref{eq:trajectory} yields~\cite{Albrecht:1997mz, Pogosian:1999np}:
\begin{equation}
    T_{00}({\bf k},\eta) \; = \; \frac{\mu\alpha}{\sqrt{1-v^2}}\frac{2\sin(k\xi\eta\cos\theta/2)}{k\cos\theta}e^{-ik({\bf x}_0\cdot\hat{\bf z}+v\eta \hat{\bf v}\cdot\hat{\bf z})} \, ,
\end{equation}
\begin{equation}
    T_{ij}({\bf k},\eta) \; = \; T_{00}({\bf k},\eta)\left[v^2 \hat{v}^i \hat{v}^j-\frac{1-v^2}{\alpha^2}\hat{X}^i \hat{X}^j\right], ~~i,j \; = \; 1,2,3 \,,
\end{equation}
where $\alpha$ is the wiggliness parameter and $\alpha=1$ corresponds to straight strings.

\subsection{UETC for global string \label{app:UETC_for_global_string}}

To calculate GW production from a network of global strings following the USM approach, one needs to account for strings with arbitrary orientations and velocities. Starting from the rest-frame expression in \eqref{eq:T00_z}, we perform the appropriate Lorentz transformations $\Lambda  = \Lambda_\phi \Lambda_\theta \Lambda_\beta $ to our static $z$-aligned string solution. Here, $\Lambda_\beta$ boosts the string in the $(\cos \psi, \sin \psi, 0)$ direction; $\Lambda_\theta$ rotates the string's orientation within the $(x,z)$ plane by a polar angle $\theta$; and $\Lambda_\phi$ governs an azimuthal rotation in the $(x,y)$ plane by angle $\phi$. As a result,

\begin{equation}
T'_{00}({\bf x}',\eta')  =  \gamma^2 T_{00}({\bf x},\eta) , \,
T'_{0i} ({\bf x}',\eta') =  \gamma^2 v_i T_{00}({\bf x},\eta), \,
T'_{ij} ({\bf x}',\eta') = \gamma^2[v_i v_j - (1-v^2) \hat{x}_i \hat{x}_j] T_{00} ({\bf x},\eta),
\label{eq:B2}
\end{equation}
where the unit vectors describing the direction of string velocity and orientation are
\begin{equation}
    \hat{\mathbf{v}} \; = \; \begin{pmatrix}
\cos\phi \cos\theta \cos\psi - \sin\phi\sin\psi\\ 
\sin\phi \cos\theta \cos\psi + \cos\phi \sin\psi\\ 
-\cos\psi\sin\theta  
\end{pmatrix}\,, ~~
\hat{\mathbf{x}} \; = \; \begin{pmatrix}
\sin\theta\cos\phi\\ 
\sin\theta\sin\phi\\ 
\cos\theta
\end{pmatrix} \, ,
\label{eq:B3}\end{equation} 
and primes indicate that we are in the boosted frame. To avoid clutter, we will drop the prime from (A.7) and onwards, with the understading that the stress tensor is evaluated in the boosted frame. Using \eqref{eq:T00_z} in conjunction with \eqref{eq:B2} allows us to calculate the Fourier Transform of the stress tensor using solely its rest-frame energy density,

\begin{equation}
\begin{aligned}
    T'_{00}(\mathbf{k},\eta') \;  &= \; \int d^3\textbf{x}\ e^{-i\textbf{k} \cdot \textbf{x}} T_{00}'(\textbf{x},\eta') \; = \; \int_{0}^{2\pi} d\phi \int_{0}^{\ell(\eta')}dz \int_{0}^\infty d\rho \ \rho^2 e^{-i\textbf{k} \cdot \textbf{x}} T_{00}'(\textbf{x},\eta') \\ 
    &=\gamma \mu(\eta') e^{-ikv\eta' \cos\psi \sin\theta + i \mathbf{k} \cdot \mathbf{x}_0 } \frac{\sin[k\ell(\eta')\cos\theta/2]}{k\cos\theta/2} \,  ,
\end{aligned}
\label{eq:B4}
\end{equation}
where we tacitly defined 
\begin{equation}
    \mu(\eta') \equiv  \pi f_a^2 \int_{\delta/a'}^{\xi \eta'} \frac{d\rho}{\rho}\ J_0(k\rho\sin\theta) \approx \pi f_a^2 \times \begin{cases}
        \ln\left[\frac{\delta^{-1}}{k/a}\right] \, , \, k\xi\eta \gg 1\\ 
        \\ 
        \ln\left[ \frac{\xi a \eta}{\delta}\right] \, , \, k\xi\eta \ll 1
    \end{cases}
\label{eq:B5}\end{equation}
and $\textbf{x}_0$ is the midpoint of the string segment. To go to the Nambu-Goto case, one simply has to omit the integral in \eqref{eq:B5}. 

Armed with \eqref{eq:B2},\eqref{eq:B4}, we can move on to calculate the UETC of two generic components of the stress tensor: 
\begin{equation}
\begin{aligned}
    T_{ij,k\ell} =& \frac{\gamma^2}{8\pi^2 V_\ell} \int_{-1}^1 d\cos\theta \int_{0}^{2\pi} d\psi\  e^{ikv(\eta-\eta')\cos\psi\sin\theta}\left[\int_0^{2\pi} d\f \left(v_i v_j - (1-v^2)\hat{x}_i \hat{x}_j\right) \times \right.\\ 
    &\times \left. \left( v_k v_\ell - (1-v^2) \hat{x}_k \hat{x}_\ell \right) \right] \times \mu(\eta) \times \mu(\eta') \times \frac{\sin\left[k\ell(\eta) \cos\theta/2\right]\sin\left[k\ell(\eta') \cos\theta/2\right]}{(k\cos\theta/2)^2}
\end{aligned}\ ,
\label{eq:B6}\end{equation}
where $V_\ell \equiv (\max(\eta,\eta'))^3$. In the $k \ell(\eta) \ll 1$ (IR) limit, equation \eqref{eq:B6} can be simplified to

\begin{equation}
\begin{aligned}
    T_{ij,k\ell} =  \frac{\mu(\eta) \mu(\eta')}{V_\ell} \cdot \frac{\gamma^2}{8\pi^2}&\int_{-1}^1 d\cos\theta \int_{0}^{2\pi} d\psi\  e^{ikv(\eta-\eta')\cos\psi\sin\theta}\left[\int_0^{2\pi} d\f \left(v_i v_j - (1-v^2)\hat{X}_i \hat{X}_j\right) \times \right.  \\ 
    &\left.\left( v_k v_\ell - (1-v^2) \hat{X}_k \hat{X}_\ell \right) \right].
    \end{aligned}
\label{eq:B7}\end{equation}

Within the USM, the correlation decays when the transverse distance of the segment is comparable to the separation of the two segments. This is equivalent to the condition $k v \vert \Delta \eta \vert \ll 1$, which means we can omit the phase factor in \cref{eq:B7} and the previous equation can be factorized into two parts: a tensorial function of velocity, and a scalar depending solely on the conformal time and wavenumber: 

\begin{equation}
    T_{ij,k\ell}= F_{ij,k\ell} \times T(k,\eta,\eta')
\label{eq:B8}\end{equation}
where 

\begin{equation}
    T(k,\eta,\eta') = \mu(\eta) \times \mu(\eta') \times \frac{\ell(\eta) \times \ell(\eta')}{(\max(\eta,\eta'))^3} \ \ , \ \ k \ell \ll 1 
\label{eq:B9}\end{equation}
with 

\begin{equation}
    F_{11,22} = 
\frac{6v^4-6v^2+1}{15(1-v^2)}     \ , \ F_{11,11} = F_{22,22} =
\frac{8v^4-8v^2+3}{15(1-v^2)}  \ , \  F_{12,12} = 
\frac{1-v^2+v^4}{15(1-v^2)} \ , \ k\ell \ll 1
\label{eq:B10}\end{equation}
and 
\begin{equation}
    \ell(\eta) = \left\{ \begin{array}{lc}
    \alpha \eta_i/2 - \Gamma_a \left( \eta - \eta_i \right)/\left(2\pi \log \left[ \frac{\xi \eta_i a_i}{\delta_c}\right] \right) \ , \ \text{Loops}\\ 
\\ 
\\ 
\xi \eta \ , \ \text{Long strings}
\end{array} \right..
\label{eq:B11}\end{equation}

By making use of \cref{eq:B9} and \cref{eq:B10} we obtain the analytical expressions in \cref{eq:T_long} and \cref{eq:T_loop}.

For completeness, we also present the $k\ell(\eta) \gg 1$ limit of the stress tensor, which is applicable only for long strings. After setting $x \equiv \cos\theta$ the triple integral becomes

\begin{equation}
    \begin{aligned}
        &T_{ij,k\ell} = \frac{\gamma^2}{4\pi^2 V_\ell} \int_{-1}^1 dx \int_{0}^{2\pi} d\psi\  e^{ikv(\eta-\eta')\cos\psi\sqrt{1-x^2}}\left[\int_0^{2\pi} d\f \left(v_i v_j - (1-v^2)\hat{x}_i \hat{x}_j\right) \times \right.\\ 
    &\times \left. \left( v_k v_\ell - (1-v^2) \hat{x}_k \hat{x}_\ell \right) \right] \times \mu(\eta) \times \mu(\eta') \times \frac{\cos(k\xi(\eta-\eta')x/2) -\cos(k\xi(\eta+\eta')x/2)  }{k^2x^2}
    \end{aligned}.
\label{eq:B12}\end{equation}

To evaluate the integral over $x$, we make use of the fact that since $k\xi \eta \gg 1$, the dominant contribution comes from the extrema of the integrand ( steepest descent method ). After some algebra, the evaluation of the integral over x simplifies \eqref{eq:B12} to 

\begin{equation}
\begin{aligned}
    T_{ij,k\ell} &= \xi \frac{\eta+\eta'-\vert \eta - \eta'\vert}{2kV_\ell} \times \mu(\eta) \times \mu(\eta') \times \frac{\gamma^2 \sqrt{\pi} e^2}{4\pi^2}\int_{0}^{2\pi} d\psi\  e^{ikv(\eta-\eta')\cos\psi}\times\\ 
    &\times \left[\int_0^{2\pi} d\f \left(v_i v_j - (1-v^2)\hat{X}_i \hat{X}_j\right) \times  \left( v_k v_\ell - (1-v^2) \hat{X}_k \hat{X}_\ell \right) \right]\Bigg|_{\cos\theta =0}\\ 
    &\equiv T^{(1)}(k,\eta,\eta') \times F_{ij,k\ell},
\end{aligned}
\label{eq:B13}\end{equation}
where 

\begin{equation}
    T(k,\eta,\eta') \equiv  \mu(\eta) \times \mu(\eta')\times \xi \times  \frac{\eta+\eta'-\vert \eta - \eta'\vert}{2k(\max(\eta,\eta'))^3} \ , \ k\xi\eta \gg 1
\label{eq:B14}\end{equation}
and 

\begin{equation}
    \frac{F_{11,22}}{\sqrt{\pi}e^2} = 
\frac{35v^4-40v^2+8}{128(1-v^2)}  \ , \ \frac{F_{11,11}}{\sqrt{\pi}e^2} = \frac{F_{22,22}}{\sqrt{\pi}e^2} =
\frac{41v^4-56v^2+24}{128(1-v^2)}  \ , \  \frac{F_{12,12}}{\sqrt{\pi}e^2} =
\frac{8-8v^2+3v^4}{128(1-v^2)} \ , \ k \xi \eta \gg 1 .
\label{eq:B15}\end{equation}

\section{Green's Functions in General Cosmological Backgrounds}
\label{app:greenfunctions}

In order to propagate the TT projection of the source to late times, we will need to find the appropriate "transfer" (Greens) functions that do that. The aim of this appendix is to derive them. 

The EOM of the tensor perturbation is, 

\begin{equation}
h''_{ij} + 2 \mathcal{H} h'_{ij} - \nabla^2 h_{ij} = 16\pi G a^2 \Pi_{ij}^{\text TT} \, . 
\label{B1}
\end{equation} 

In Fourier space this reads 

\begin{equation}
h''_{ij} + 2 \mathcal{H} h'_{ij} + k^2 h_{ij}  =16 \pi G a^2 \Pi_{ij}^{\text TT} . 
\label{B2}
\end{equation}

In order to solve this equation, we shall make use of the retarded Greens function. This function satisfies 

\begin{equation}
\mathcal{G}'' + 2 \mathcal{H} \mathcal{G}' + k^2 \mathcal{G} = \delta(\eta - \eta'). 
\label{B3}
\end{equation}

The explicit form of the Green's function depends on the cosmological era through the evolution of $\mathcal{H}$. For power-law expansion $a(\eta) \propto \eta^\alpha$, we have $\mathcal{H} = \alpha/\eta$, and the homogeneous equation admits analytical solutions. For general cosmological histories, the general solution in terms of two independent solutions $u(x)$ and $v(x)$ of the homogeneous equation that satisfies the boundary condition is
\begin{equation}
\mathcal{G}(x,y) \; = \; \Theta(x-y) \frac{u(y)v(x) - u(x)v(y)}{W[u,v](y)} \, ,
\label{eq:greenexplicit}
\end{equation}
where  $W[u,v] = u w' - u' v$ is the Wronskian.

For the case of a power law scale factor, $a \propto \eta^\alpha$, we can rewrite equation \eqref{B3} in terms of the dimensionless variable $ x \equiv k\eta$: 

\begin{equation}
\frac{d^2\mathcal{G}}{dx^2} + \frac{2\alpha}{x} \frac{d\mathcal{G}}{dx} + \mathcal{G} = \frac{1}{k} \delta(x-y). 
\label{B4}
\end{equation}
We will first solve the homogeneous equation. We find that there are two solutions,
\begin{equation}
    \begin{aligned}
        u(x) \; = \; c x^{\frac{1}{2} - \alpha} J_{\alpha - \frac{1}{2}}(x) \, , \\
        v(x) \; = \; c x^{\frac{1}{2} - \alpha} Y_{\alpha - \frac{1}{2}}(x) \, , 
    \end{aligned}
\end{equation}
where $c$ is an arbitrary constants and $J_\nu(x), Y_\nu(x)$ are the Bessel functions of the first and second kind, respectively. If we use the general expression for the Green's function solution~\eqref{eq:greenexplicit}, we find the solution
\begin{equation}
\label{eq:RDGreensFunction}
    \mathcal{G}(x, y) \; = \; \frac{\pi}{2} \Theta(x - y) x^{\frac{1}{2} - \alpha} y^{\frac{1}{2} + \alpha} \left[J_{\alpha - \frac{1}{2}}(y) Y_{\alpha - \frac{1}{2}}(x) - J_{\alpha - \frac{1}{2}}(x) Y_{\alpha - \frac{1}{2}}(y) \right] \, .
\end{equation}
For the radiation background, with $\alpha = 1$, we find
\begin{equation}
        \mathcal{G}(x, y) \; = \; \Theta(x - y) \cdot \frac{y \sin(x-y)}{x} ~(\rm{RD})\, , 
\end{equation}
and for matter dominated background, with $\alpha = 2$, we recover
\begin{equation}
\label{eq:MDGreensFunction}
    \mathcal{G}(x, y) \; = \; \Theta(x - y) \cdot \left[\frac{y(y-x)\cos(x-y) + y(1+xy)\sin(x-y)}{x^3} \right]~(\rm{MD}) \, .
\end{equation}

\section{Mode Matching}
\label{app:ModeMatching}

The aim of this appendix is to explain the propagation of wavemodes throughout the history and after the collapse of the network. 

The behavior of cosmological perturbations can be split into two categories: those that are inside the horizon and obey $k\eta \gg 1$ (sub-horizon), and those that are outside the horizon and obey $k \eta \ll 1$ (super-horizon). Depending on when (and if) horizon re-entry happens, the evolution of the corresponding mode gets altered significantly. Our interest lies within the following scenarios: 

\begin{enumerate}
    \item[]\textit{Case I:} Modes that are super-horizon when the network collapses in RD and re-enter in MD.
    \item[]\textit{Case II:} Modes that are super-horizon when the network collapses during MD and they do not re-enter.
    \item[]\textit{Case III:} Modes that are sub-horizon when the network collapses during MD.
    \item[]\textit{Case IV:} Modes that are sub-horizon, with the network collapsing at equality between RD and MD.
\end{enumerate}

In principle, the Greens functions solution quoted in \cref{eq:OmGW_master} encompasses any and all effects that may arise in each of these circumstances, with $G(u_0,u)$ incorporating the details of sub and super-horizon physics as well as of the different eras. Its generic form can be found by solving \cref{B3} for a scale factor that describes matter and radiation jointly,

\begin{equation}
    a(t) = a_0^2 H_0 \sqrt{\Omega_r} \left( \frac{g_{s}(t)}{g_{s,0}} \right)^{-1/6} \eta + \frac{a_0^2}{4} H_0^2 \Omega_m \eta^2  .
\label{eq:genericscalefactor}\end{equation}

The solutions for $\Omega_m=0$ and $\sqrt{\Omega_r} =0$ were found in \cref{eq:RDGreensFunction} and \cref{eq:MDGreensFunction}, respectively. For a generic linear combination of the two no analytical solution is known to date. To alleviate this, we approximate the background transition from RD to MD by an instantaneous transition at $\eta_{\rm eq}$. Under this approximation, we shall consider the evolution of the modes in each era separately and invoke the continuity of the GW amplitude and its derivative to propagate them across the different eras. To facilitate the discussion, we are going to define a dimensionless ``spectral function" $\mathcal{I}$ as 

\begin{equation}
\mathcal{I}(w,w_f) \equiv \frac{\xi}{k(\mu_{\rm eff}(\eta_f))^2} \int_{w_i}^{w_f} dw' \int_{w_i}^{w_f} dw''\ G(w,w') G(w,w'') T(k,\frac{w'}{k},\frac{w''}{k}).
\label{eq:DimSpecFuncDefinition}\end{equation}

The full gravitational wave intensity is then given by evaluating the spectral function at the present time:

\begin{equation}
    \Omega_{GW}(k) = \frac{128 (G \mu_{\rm eff}(\eta_f))^2 k^2}{3a_0^2 H_0^2} \cdot F_{12,12} \cdot \xi^{-1} \cdot \mathcal{I}(w_0,w_f).
\label{eq:OmegaGWwithDimSpecFunc}\end{equation}

We will analyze global and gauge strings jointly since the results of the latter can be informed from the former by simply removing the scale dependent logarithm from $\mu: \mu_{\rm eff}(\eta_f) \to \pi f_a^2$.
Starting with the RD case, i.e. $w,w_f < w_{\rm eq}$ (with $\eta_{\rm eq} = \frac{w_{\rm eq}}{k}$ corresponding to equality between MD and RD) we find that we can analytically compute $\mathcal{I}$: 

\begin{equation}
\begin{aligned} 
    \mathcal{I}^{\rm RD}(w,w_f) &=\frac{1}{w_0^2} \int_{w_i}^{w_f} dw' \int_{w_i}^{w_f} dw''\ \frac{(w'w'')^2}{(\max(w',w''))^3} \sin(w-w') \sin(w-w'')\\ 
    &= \frac{2}{w_0^2} \int_{w_i}^{w_f}\frac{dw'}{w'} \sin(w-w') \int_{w_i}^{w'}dw''\ (w'')^2 \cdot \sin(w-w'')
\end{aligned}.
\label{eq:RDdimspecstep1}\end{equation}

Evaluating the two integrals yields 

\begin{equation}
\begin{aligned}
    \mathcal{I}^{RD}(w,w_f) = \frac{2}{w^2} &\left\{ \int_{w_f}^{2w_f} \frac{1-\cos(t)}{t} dt+ w_f+ \frac{w_f}{4} \cos(2(w-w_f))- \frac{5}{8}\sin(2w) \right.\\ 
    &\left.+\frac{5}{8}\sin(2(w-w_f)) - 2\cos^2(w) {\rm Si}(w_f) + \cos(2w) {\rm Si}(2w_f) \right\}
\end{aligned},
\label{eq:RDdimspecfinal}\end{equation}
where $Si$ is the integral sine function. The previous expression can be expanded around $w_f$ to give 

\begin{equation}
\mathcal{I}^{\rm RD}(w,w_f) = \frac{2}{9} \left( \frac{\sin(w)}{w} \right)^2 w_f^3\left(1 + \mathcal{O}(w_f)^2\right).
\label{eq:IRD}\end{equation}

In a similar fashion one can perform the exact same calculation for MD, albeit more cumbersome: 

\begin{equation}
\begin{aligned}
\mathcal{I}^{\rm MD}(w,w_f) = \frac{1}{8w^6} \Big[ 
& -64\,\mathrm{Ci}(w_f)\Big((w^2-1)\sin(2w)+2w\cos(2w)\Big) \\
& +64\,\mathrm{Ci}(2w_f)\Big((w^2-1)\sin(2w)+2w\cos(2w)\Big) \\
& -64\,w^2\,\mathrm{Si}(w_f)
+64\,w^2\,\mathrm{Si}(w_f)\cos(2w)
-64\,w^2\,\mathrm{Si}(2w_f)\cos(2w) \\
& -128\,w\,\mathrm{Si}(w_f)\sin(2w)
+128\,w\,\mathrm{Si}(2w_f)\sin(2w) \\
& -64\,\mathrm{Si}(w_f)\cos(2w)
+64\,\mathrm{Si}(2w_f)\cos(2w)
-64\,\mathrm{Si}(w_f) \\
& +8\,w^2 w_f^3
+4\,w^2 w_f^3 \cos\!\big(2(w-w_f)\big)
+26\,w^2 w_f^2 \sin\!\big(2(w-w_f)\big) \\
& -101\,w^2 \sin\!\big(2(w-w_f)\big)
+64\,w^2 \sin(2w-w_f)
+64\,w^2 \sin(w_f) \\
& -74\,w^2 w_f \cos\!\big(2(w-w_f)\big)
+37\,w^2 \sin(2w)
-64\,w^2 \log(2)\sin(2w) \\
& -8\,w w_f^3 \sin\!\big(2(w-w_f)\big)
-4\,w_f^3 \cos\!\big(2(w-w_f)\big)
-26\,w_f^2 \sin\!\big(2(w-w_f)\big) \\
& +52\,w w_f^2 \cos\!\big(2(w-w_f)\big)
+148\,w w_f \sin\!\big(2(w-w_f)\big)
+101\,\sin\!\big(2(w-w_f)\big) \\
& -64\,\sin(2w-w_f)
+74\,w_f \cos\!\big(2(w-w_f)\big)
-202\,w \cos\!\big(2(w-w_f)\big) \\
& +128\,w \cos(2w-w_f)
-37\,\sin(2w)
+74\,w \cos(2w)
+64\,\log(2)\sin(2w) \\
& -128\,w \log(2)\cos(2w)
+8\,w_f^3
+64\,\sin(w_f)
\Big].
\end{aligned}
\label{eq:IMDstep1}\end{equation}

Armed with the general expressions for the dimensionless spectral function, we can go ahead and discuss cases \textit{(I)- (IV)}: 

\subsection*{Case I:}

The gravitational wave amplitude $h$ generated by the collapsing network stays constant up until re-entry. That is, after it re-enters the horizon it redshifts as free radiation. For re-entry at MD we have

\begin{equation}
    h(\eta_0,k) \approx -3 h(\eta_f,k) \cdot \frac{\cos(k\eta_0)}{(k\eta_0)^2} \implies \vert h_0 \vert^2 \approx \frac{9}{2 (k\eta_0)^4} \vert h_f \vert^2 .  
\label{eq:hredshiftMD}\end{equation}

This, in turn, means that the dimensionless spectral function $\mathcal{I}(w_0,w_f)$ is 

\begin{equation}
    \mathcal{I}(w_0,w_f) = \frac{9}{2 w_0^4} \mathcal{I}^{\rm RD}(w_{\rm eq},w_f) \approx \frac{w_f^3}{2w_{0}^4}, 
\label{eq:dimspeccaseI}\end{equation}
where we tacitly assumed that $w_{\rm eq} \ll 1$. For the case of re-entry at RD we have 

\begin{equation}
    h(\eta_0,k) \approx h(\eta_f,k) \cdot \frac{\sin(k\eta_0)}{k\eta_0} \implies \vert h_0 \vert^2 \approx \frac{1}{2(k\eta_0)^2} \vert h_f \vert^2 
\end{equation}
and in turn this yields 

\begin{equation}
    \mathcal{I}(w_0,w_f) = \frac{1}{2w_0^2} \mathcal{I}^{\rm RD}(w_{\rm eq},w_f) \approx \frac{1}{9} \frac{w_f^3}{w_0^2}. 
\end{equation}

 Plugging these expressions into \cref{eq:OmegaGWwithDimSpecFunc} reproduces the expressions in the main text.
 
\subsection*{Case II:} 

Modes that are super-horizon during MD have not re-entered the horizon today and froze in when the network collapsed. The dimensionless spectral function today can be approximated as, 

\begin{equation}
    \mathcal{I}(w_0,w_f) = \mathcal{I}^{\rm MD}(w_f,w_f) = \frac{w_f^3}{18}.
\label{eq:dimspeccaseII}\end{equation}

This reproduces the GW amplitude quoted in \cref{eq:Omega_gw_I}. 

\subsection*{Case III:} 

For modes that obey $1 < w_f < \xi^{-1}$, the formula \cref{eq:IMDstep1} can be found to have the following simple expression:

\begin{equation}
    \mathcal{I}^{\rm MD}(w_0,w_f) \approx \frac{1}{2} \frac{w_f^5}{w_0^4}.
\label{eq:IMDsub}\end{equation}

Upon substituting into \cref{eq:OmegaGWwithDimSpecFunc} we obtain \cref{eq:Omega_gw_II}.

\subsection*{Case IV:} 

Finally, for the case of a sub-horizon mode in RD,  the expression \cref{eq:IRD} reproduces the correct amplitude remarkably well even though it was obtained by expanding around $w_f = 0$. Setting $w = w_0$ yields the following: 

\begin{equation}
    \mathcal{I}^{\rm RD}(w_0,w_f) \approx \frac{1}{9} \frac{w_f^3}{w_0^2}.
\label{eq:IRDsub}\end{equation}

Upon substituting into \cref{eq:OmegaGWwithDimSpecFunc} we obtain \cref{eq:Omega_gw_II}.

\newpage 

\addcontentsline{toc}{section}{References}
\bibliographystyle{utphys}
\bibliography{references} 

\end{document}